\documentclass[11pt,reqno]{amsart}

\usepackage{amsmath}
\usepackage{amssymb}
\usepackage{a4wide}
\usepackage{bbm}
\usepackage{graphics}
\usepackage{color}
\usepackage[applemac]{inputenc}
\usepackage[sans]{dsfont}

\newtheorem{theorem}{Theorem}[section]
\newtheorem{lemma}[theorem]{Lemma}

\newtheorem{remarks}[theorem]{Remarks}

\numberwithin{equation}{section}

\renewcommand{\d}{\mathrm{d}}
\newcommand{\el}{\mathrm{el}}
\newcommand{\fib}{\mathrm{fib}}
\newcommand{\n}{\mathrm{n}}
\newcommand{\ph}{\mathrm{ph}}
\renewcommand{\i}{\mathrm{i}}
\newcommand{\DETAILS}[1]{}

\begin{document}

\title[Uniqueness of the dressed hydrogen atom ground state]{Hyperfine splitting in non-relativistic QED: uniqueness of the dressed hydrogen atom ground state}

\author[L. Amour]{Laurent Amour}
\address[L. Amour]{Laboratoire de Math\'ematiques \\ Universit\'e de Reims \\ Moulin de la Housse, BP 1039, 51687 REIMS Cedex 2, France, et FR-CNRS 3399}
\email{laurent.amour@univ-reims.fr}
\author[J. Faupin]{J{\'e}r{\'e}my Faupin}
\address[J. Faupin]{Institut de Math{\'e}matiques de Bordeaux \\
UMR-CNRS 5251, Universit{\'e} de Bordeaux 1 \\
351 cours de la lib{\'e}ration, 33405 Talence Cedex, France}
\email{jeremy.faupin@math.u-bordeaux1.fr}

\date{}

\begin{abstract}
We consider a free hydrogen atom composed of a spin-$\frac 12$ nucleus and a spin-$\frac 12$ electron in the standard model of non-relativistic QED. We study the Pauli-Fierz Hamiltonian associated with this system at a fixed total momentum. For small enough values of the fine-structure constant, we prove that the ground state is unique. This result reflects the hyperfine structure of the hydrogen atom ground state.
\end{abstract}

\maketitle

\section{Introduction}

The structure of the spectrum of the Pauli Hamiltonian describing a non-relativistic Hydrogen atom in Quantum Mechanics is well-known (see e.g. \cite{CTDL}). Among many properties, a remarkable one is that the interaction between the spins of the nucleus and the electron causes the so-called \emph{hyperfine} structure of Hydrogen. In particular, in spite of the spins degrees of freedom, the ground state of the Pauli Hamiltonian is unique; A three-fold degenerate eigenvalue appears besides, close to the ground state energy. This phenomenon justifies the famous observed \emph{$21$-cm Hydrogen line}. Mathematically, using standard perturbation theory of isolated eigenvalues, this statement is not difficult to establish.

In the framework of non-relativistic QED, due to the absence of mass of photons, the bottom of the spectrum of the Hamiltonian coincides with the bottom of its essential spectrum. Thus the question of the (existence and) multiplicity of the ground state is much more subtle.

In this paper we consider a moving Hydrogen atom in non-relativistic QED. The total system (electron, nucleus and photons) is translation invariant, hence one can fix the total momentum and study the corresponding fiber Hamiltonian. For sufficiently small values of the total momentum, the bottom of the spectrum is known to be an eigenvalue (see \cite{AGG2,LMS1}). Moreover, under simplifying assumptions, the multiplicity of the ground state eigenvalue is also known: If both the electron and nucleus spins are neglected, the ground state is unique \cite{AGG2}. If the electron spin is taken into account and the nucleus spin is neglected, then the ground state is twice degenerate \cite{LMS2}.

Now, following the physical prescription, we assume that both the electron and the nucleus have a spin equal to $\frac 12$. In \cite{AF2}, using a contradiction argument, we proved under this assumption that the multiplicity of the ground state is strictly less than $4$. This shows that a hyperfine splitting does occur in non-relativistic QED (we refer the reader to the introduction of \cite{AF2} for a more detailed discussion on the hyperfine structure of Hydrogen). Refining our previous analysis, we shall prove in the present paper that the ground state of the dressed Hydrogen atom is unique.

\subsection{Definition of the model and main results}

We recall the definition of the Hamiltonian associated with a freely moving hydrogen atom at a fixed total momentum $P$ in non-relativistic QED. For more details, we refer to \cite[Section 2]{AF2}.

The Hilbert space of the total system is
\begin{equation*}
\mathcal{H}_{\fib} := \mathbb{C}^4 \otimes \mathrm{L}^2( \mathbb{R}^3 ) \otimes \mathcal{H}_{\mathrm{ph}}, \quad \text{where} \quad \mathcal{H}_{\mathrm{ph}} := \mathbb{C} \oplus \bigoplus_{n=1}^\infty S_n \left [ \mathrm{L}^2( \mathbb{R}^3 \times \{1,2\} )^{ \otimes^n} \right ].
\end{equation*}
The photon space $\mathcal{H}_\ph$ is the symmetric Fock space over $\mathrm{L}^2( \mathbb{R}^3 \times \{1,2\} )$ ($S_n$ denotes the symmetrization operator). In units such that the Planck constant divided by $2\pi$ and the velocity of light are equal to $1$, the Hamiltonian we consider acts on $\mathcal{H}_{\mathrm{fib}}$ and is given by the expression
\begin{align}
 H_g( P ) =& \frac{1}{2m_\el} \left( \frac{ m_\el }{ M } ( P - P_\ph ) + p_r - g A( \frac{ m_\el }{ M } g^{\frac{2}{3}} r )\right )^2 \notag \\
 & + \frac{1}{2m_\n} \left( \frac{ m_\n }{ M } ( P - P_\ph ) - p_r + g A( - \frac{ m_\n }{M} g^{\frac{2}{3}} r ) \right)^2 \notag \\
& - \frac{ 1 }{ |r| } + H_{\rm ph}   -\frac{g}{2m_\el} \sigma^{\el} \cdot B( \frac{ m_\el }{ M } g^{\frac{2}{3}} r )
   + \frac{g}{2m_\n} \sigma^{\n} \cdot B ( - \frac{ m_\n }{ M }  g^{\frac{2}{3}} r ). \label{eq:def_Hg(P)}
\end{align}
The coupling parameter $g$ is given by $g:=\alpha^{ 3/2 }$ where $\alpha=e^2$ is the fine-structure constant (with $e$ the charge of the electron). The variable $r$ is the intern position variable of the hydrogen atom, and $p_r := - \i \nabla_r$ is the associated momentum operator. The masses $m_\el$, $m_\n$ and $M := m_\el + m_\n$ are respectively the mass of the electron, the mass of the nucleus and the total mass of the atom. The $3$-uples $\sigma^\el = ( \sigma^\el_1 , \sigma^\el_2 , \sigma^\el_3 )$ and $\sigma^\n = ( \sigma^\n_1 , \sigma^\n_2 , \sigma^\n_3 )$ are the Pauli matrices associated with the spins of the electron and the nucleus respectively, that is
\begin{footnotesize}
\begin{align*}
\sigma^\el_1 &=
  \left (
  \begin{array}{cccc}
   0 & 0 & 1 & 0 \\
   0 & 0 & 0 & 1 \\
   1 & 0 & 0 & 0 \\
   0 & 1 & 0 & 0
  \end{array}
  \right ) , \quad
  \sigma^\el_2 =
    \left (
  \begin{array}{cccc}
   0 & 0 & - \i & 0 \\
   0 & 0 & 0 & -\i \\
   \i & 0 & 0 & 0 \\
   0 & \i & 0 & 0
  \end{array}
  \right ) , \quad
  \sigma^\el_3 =
    \left (
  \begin{array}{cccc}
   1 & 0 & 0 & 0 \\
   0 & 1 & 0 & 0 \\
   0 & 0 & -1 & 0 \\
   0 & 0  & 0 & -1
  \end{array}
  \right ) ,  \\
 \sigma^\n_1 &=
  \left (
  \begin{array}{cccc}
   0 & 1 & 0 & 0 \\
   1 & 0 & 0 & 0 \\
   0 & 0 & 0 & 1 \\
   0 & 0 & 1 & 0
  \end{array}
  \right ) , \quad
  \sigma^\n_2 =
    \left (
  \begin{array}{cccc}
   0 & -\i & 0 & 0 \\
   \i & 0 & 0 & 0 \\
   0 & 0 & 0 & -\i \\
   0 & 0 & \i & 0
  \end{array}
  \right ) , \quad
  \sigma^\n_3 =
    \left (
  \begin{array}{cccc}
   1 & 0 & 0 & 0 \\
   0 & -1 & 0 & 0 \\
   0 & 0 & 1 & 0 \\
   0 & 0  & 0 & -1
  \end{array}
  \right ). 
\end{align*}
\end{footnotesize}

As usual, for any $h \in \mathrm{L}^2( \mathbb{R}^3 \times \{1,2\} )$, we set
\begin{equation*}
a^*(h) := \sum_{\lambda=1,2} \int_{\mathbb{R}^3} h( k,\lambda ) a^*_\lambda(k) \d k, \quad a(h) := \sum_{\lambda=1,2} \int_{\mathbb{R}^3} \bar h( k,\lambda ) a_\lambda( k) \d k,
\end{equation*}
and $\Phi(h) := a^*(h) + a(h)$,  where the creation and annihilation operators, $a^*_\lambda(k)$ and $a_\lambda(k)$, are operator-valued distributions obeying the canonical commutation relations
\begin{equation*}
[ a_\lambda (k) , a_{\lambda'}(k') ] = [ a^*_\lambda(k) , a^*_{\lambda'}(k') ] = 0 , \quad [ a_\lambda ( k ) , a^*_{\lambda'}( k' ) ] = \delta_{\lambda\lambda'} \delta ( k - k' ).
\end{equation*}
For $x \in \mathbb{R}^3$, the vectors $A(x)$ and $B(x)$ of the quantized electromagnetic field in the Coulomb gauge are defined by
\begin{align}
& A(x) := \frac{1}{2\pi} \sum_{\lambda =1,2} \int_{ \mathbb{R}^3 } \frac{ \chi_\Lambda (k) }{ |k|^{\frac{1}{2}} } \varepsilon^\lambda(k) \left [ e^{ - \i k \cdot x } a^*_{\lambda }(k) + e^{ \i k \cdot x}a_{\lambda} (k)\right ] \d k, \label{eq:defA} \\
& B(x) := -\frac{\i}{2\pi} \sum_{\lambda =1,2} \int_{ \mathbb{R}^3 } |k|^{\frac{1}{2} } \chi_\Lambda(k) \big( \frac{k}{|k|} \wedge \varepsilon^{\lambda}(k) \big) \left [ e^{ - \i k \cdot x} a^*_{\lambda} (k) - e^{ \i k \cdot x}a_{\lambda} (k)\right ] \d k, \label{eq:defB}
\end{align}
where the polarization vectors are chosen in the following way:
\begin{equation*}
\varepsilon^1(k): = \frac{  ( k_2 , -k_1 , 0 ) }{ \sqrt{ k_1^2 + k_2^2 } } \quad , \quad \varepsilon^2(k) := \frac{ k }{ |k| } \wedge \varepsilon^1 (k) = \frac{ ( - k_1 k_3 , - k_2 k_3 , k_1^2 + k_2^2 ) }{ \sqrt{ k_1^2 + k_2^2 } \sqrt{ k_1^2 + k_2^2 + k_3^2 } }.
\end{equation*}
In particular, for $j \in \{ 1,2,3 \}$, we have $A_j(x) = \Phi( h^A_j(x) )$
and $B_j(x) = \Phi( h^B_j(x) )$,
with
\begin{align}
& h^A_j(x,k,\lambda) := \frac{1}{2\pi} \frac{ \chi_\Lambda(k) }{ |k|^{ \frac{1}{2} } } \varepsilon^\lambda_j(k) e^{ - \i k \cdot x }, \label{eq:h^A} \\
& h^B_j(x,k,\lambda) := - \frac{\i}{2\pi} |k|^{\frac{1}{2}} \chi_\Lambda(k) \left ( \frac{ k }{ |k| } \wedge \varepsilon^\lambda(k) \right )_j e^{ - \i k \cdot x }. \label{eq:h^B}
\end{align}
In \eqref{eq:defA} and \eqref{eq:defB}, $\chi_\Lambda(k)$ denotes an ultraviolet cutoff function which is chosen, for simplicity, as $\chi_\Lambda(k) := \mathds{1}_{ |k| \le \Lambda }(k)$. Here, $\Lambda$ is supposed to be a given arbitrary (large and) positive parameter.

The Hamiltonian and total momentum of the free photon field, $H_\ph$ and $P_\ph$, are defined by
\begin{equation*}
H_{\mathrm{ph}} := \sum_{ \lambda = 1, 2 } \int_{ \mathbb{R}^3 } |k| a^*_\lambda(k) a_\lambda(k) \d k, \qquad P_\ph := \sum_{\lambda = 1,2} \int_{ \mathbb{R}^3 } k a^*_\lambda(k) a_\lambda(k) \d k.
\end{equation*}
The Fock vacuum is denoted by $\Omega$.

Our main result is the following.
\begin{theorem}\label{thm:main}
There exist $g_c>0$ and $p_c>0$ such that, for all $0<g\le g_c$ and $0 \le |P| \le p_c$, $H_g(P)$ has a unique ground state, that is
\begin{align*}
E_g(P) := \inf \mathrm{spec} (H_g(P)) \text{ is a simple eigenvalue of } H_g(P).
\end{align*}
\end{theorem}
\begin{remarks}\label{rk:wick}
$\quad$
\begin{itemize}
\item[(1)] For convenience, we shall work in the sequel with the Hamiltonian \emph{:}$H_g(P)$\emph{:} obtained from the expression \eqref{eq:def_Hg(P)} by Wick ordering. Since \emph{:}$H_g(P)$\emph{:} and $H_g(P)$ only differ by a constant, the statement of Theorem \ref{thm:main} is equivalent if we replace \emph{:}$H_g(P)$\emph{:} by $H_g(P)$. From now on, to keep notations simple, we use $H_g(P)$ to designate the Wick-ordered Hamiltonian.
\item[(2)] With some more work, Theorem \ref{thm:main} may be proven with the critical value $p_c = M  - \varepsilon$, $\varepsilon > 0$, and $g_c$ depending on $\varepsilon$. However, for large values of the total momentum, $|P| > M$, due to Cerenkov radiation, one expects that $E_g(P)$ is not an eigenvalue.
\end{itemize}
\end{remarks}

\subsection{Notations and strategy of the proof}

Now we describe the strategy of our proof and introduce corresponding notations. Our main tools will be a suitable \emph{infrared decomposition} of Fock space combined with \emph{iterative perturbation theory}. The introduction of an infrared cutoff into the interaction Hamiltonian is a standard step in the analysis of models of non-relativistic QED \cite{Fr}. The idea of considering a sequence of Hamiltonians with decreasing infrared cutoffs and comparing them iteratively by perturbation theory can be traced back to \cite{Pizzo}. It was later used successfully in different contexts \cite{BFP,CFP,AFGG2,BG}.

Roughly speaking, the method employed in these papers to prove the existence of a unique ground state is as follows: Let $H_\sigma$ denote the Hamiltonian with an infrared cutoff of parameter $\sigma$, acting on the Fock space of particles of energies $\ge\sigma$. For large $\sigma$'s, there is no interaction in $H_\sigma$ and it is easy to verify that $H_\sigma$ has a unique ground state separated by a gap of order $\mathcal{O}(\sigma)$ from the rest of the spectrum. Next, using perturbation theory, one shows that, if for some given $\sigma>0$, $H_\sigma$ fulfills this gap property (uniqueness of the ground state and gap of order $\mathcal{O}(\sigma)$ above it), then the same holds for $H_{\sigma/2}$. Proceeding iteratively, one thus obtains the existence of a unique ground state for any $\sigma>0$. Finally a suitable use of a `pull-through' argument implies the result for $\sigma=0$.

In all the previously cited papers, the ground state of the non-interacting Hamiltonian is unique. In our context, however, it is $4$-fold degenerate, so that the method does not directly apply. Let us be more precise. For $\sigma>0$, the infrared (fibered) Hamiltonian acts on $\mathcal{H}_\fib$ and is defined as
\begin{align}
 H_{g,\ge\sigma}( P ) =& \frac{1}{2m_\el} \text{:} \left( \frac{ m_\el }{ M } ( P - P_\ph ) + p_r - g A_{\ge\sigma}( \frac{ m_\el }{ M } g^{\frac{2}{3}} r )\right )^2 \text{:} \notag \\
 & + \frac{1}{2m_\n} \text{:} \left( \frac{ m_\n }{ M } ( P - P_\ph ) - p_r + g A_{\ge\sigma}( - \frac{ m_\n }{M} g^{\frac{2}{3}} r ) \right)^2 \text{:} \notag \\
& - \frac{ 1 }{ |r| } + H_{\rm ph}   -\frac{g}{2m_\el} \sigma^{\el} \cdot B_{\ge\sigma}( \frac{ m_\el }{ M } g^{\frac{2}{3}} r )
   + \frac{g}{2m_\n} \sigma^{\n} \cdot B_{\ge\sigma} ( - \frac{ m_\n }{ M }  g^{\frac{2}{3}} r ), \label{eq:def_Hgsigma}
\end{align}
where, for any $j\in\{1,2,3\}$, $x \in \mathbb{R}^3$, $k \in \mathbb{R}^3$, $\lambda \in \{1,2\}$ and $\sigma \ge 0$,
\begin{align*}
& A_{j,\ge\sigma}(x) := \Phi ( h_{j,\ge\sigma}^A(x) ) ~ \text{ with } ~ h_{j,\ge\sigma}^A(x,k,\lambda) := \mathds{1}_{ |k| \ge \sigma }(k) h_j^A(x,k,\lambda), \\
& B_{j,\ge\sigma}(x) := \Phi ( h_{j,\ge\sigma}^B(x) ) ~ \text{ with } ~ h_{j,\ge\sigma}^B(x,k,\lambda) := \mathds{1}_{ |k| \ge \sigma }(k) h_j^B(x,k,\lambda).
\end{align*}
The expression \eqref{eq:def_Hgsigma} is Wick ordered in accordance with Remark \ref{rk:wick} (1).
For $\sigma>0$, let
\begin{align*}
& \mathcal{H}_{\ph,\ge \sigma} := \mathbb{C} \oplus \bigoplus_{n=1}^\infty S_n \left [ \mathrm{L}^2( \{ (k,\lambda) \in \mathbb{R}^3 \times \{1,2\} , |k| \ge \sigma \} )^{ \otimes^n} \right ], \\
& \mathcal{H}_{\ph,\le \sigma} := \mathbb{C} \oplus \bigoplus_{n=1}^\infty S_n \left [ \mathrm{L}^2( \{ (k,\lambda) \in \mathbb{R}^3 \times \{1,2\} , |k| \le \sigma \} )^{ \otimes^n} \right ],
\end{align*}
denote the Fock spaces for photons of energies $\ge \sigma$, respectively of energies $\le \sigma$. It is well-known that there exists a unitary transformation mapping $\mathcal{H}_{\ph}$ to $\mathcal{H}_{\ph,\ge\sigma} \otimes \mathcal{H}_{\ph,\le\sigma}$. 
\begin{equation*}
\mathcal{H}_{\fib,\ge\sigma} := \mathbb{C}^4 \otimes \mathrm{L}^2( \mathbb{R}^3 , \d r ) \otimes \mathcal{H}_{\ph,\ge\sigma}.
\end{equation*}
Clearly, $\mathcal{H}_{\fib,\ge\sigma}$ identifies with a subset of $\mathcal{H}_{\fib}$, and
\begin{equation*}
H_{g,\ge\sigma}(P) : \mathcal{H}_{\fib,\ge\sigma} \cap D( H_{g,\ge\sigma}(P) ) \to \mathcal{H}_{\fib,\ge\sigma}.
\end{equation*}
The restriction of $H_{g,\ge\sigma}(P)$ to $\mathcal{H}_{\fib,\ge\sigma} \cap D( H_{g,\ge\sigma}(P) )$ is then denoted by
\begin{equation*}
K_{g, \ge \sigma} (P) := H_{g,\ge\sigma}(P) |_{ \mathcal{H}_{\fib,\ge\sigma} \cap D( H_{g,\ge\sigma}(P) ) }.
\end{equation*}
In order to avoid any confusion, we also set
\begin{align*}
& H_{\ph, \ge \sigma} := H_{\ph} |_{ \mathcal{H}_{\ph,\ge\sigma}  \cap D( H_\ph ) }, \quad P_{\ph, \ge \sigma} := P_{\ph} |_{ \mathcal{H}_{\ph,\ge\sigma}  \cap D( P_\ph ) },
\end{align*}
and the vacuum in $\mathcal{H}_{\ph,\ge\sigma}$ is denoted by $\Omega_{\ge\sigma}$. We shall use the decomposition
\begin{equation*}
K_{g,\ge\sigma}(P) = K_{0,\ge\sigma}(P) + W_{g,\ge\sigma}(P),
\end{equation*}
where
\begin{align}
K_{0,\ge\sigma}(P) :=& H_0(P) |_{ D(H_0(P)) \cap \mathcal{H}_{\fib,\ge\sigma} } = H_r + \frac{1}{2M} ( P - P_{\ph,\ge\sigma} )^2 + H_{ \ph,\ge\sigma }, \label{eq:def_K0}
\end{align}
and
\begin{align}
W_{g,\ge\sigma}(P) =& - \frac{ g }{ m_\el } \left ( \big ( \frac{ m_\el }{ M } ( P - P_{\ph,\ge\sigma} ) + p_r \big ) \cdot A_{\ge\sigma} ( \frac{ m_\el }{ M } g^{\frac{2}{3}} r ) \right ) \notag \\ 
& + \frac{ g }{ m_\n } \left ( \big ( \frac{ m_\n }{ M } ( P - P_{\ph,\ge\sigma} ) - p_r \big ) \cdot A_{\ge\sigma} ( - \frac{ m_\n }{ M } g^{\frac{2}{3}} r ) \right ) \notag \\ 
& + \frac{ g^2 }{ 2m_\el } \text{:} A_{\ge\sigma} ( \frac{ m_\el }{ M } g^{\frac{2}{3}} r )^2\text{:} + \frac{ g^2 }{ 2m_\n } \text{:} A_{\ge\sigma} ( - \frac{ m_\n }{ M } g^{\frac{2}{3}} r )^2 \text{:} \notag \\
& -\frac{g}{2m_\el} \sigma^{\el} \cdot B_{\ge\sigma}( \frac{ m_\el }{ M } g^{\frac{2}{3}} r ) + \frac{g}{2m_\n} \sigma^{\n} \cdot B_{\ge\sigma} ( - \frac{ m_\n }{ M } g^{\frac{2}{3}} r ). \label{eq:Wg>sigma}
\end{align}
In \eqref{eq:def_K0}, $H_r$ denotes the Schr{\"o}dinger Hamiltonian
\begin{align*}
H_r:= \frac{p^2_r}{2\mu} - \frac{1}{ |r| },
\end{align*}
where $\mu$ is the reduced mass, $\mu := m_1m_2/(m_1+m_2)$.

For any self-adjoint and semi-bounded operator $H$, we set 
$E(H) := \inf \mathrm{spec}  (H)$ and
\begin{align*}
\mathrm{Gap}( H ) := \inf ( \mathrm{spec} ( H ) \setminus \{ E(H) \} ) - E(H).
\end{align*}
We then  observe that, for all $\sigma \ge 0$,
\begin{align*}
& E( K_{0,\ge\sigma}(P) ) = E( H_r ) + \frac{P^2}{2M} =: e_0 + \frac{P^2}{2M} =: E_0(P), 
\end{align*}
and that $E_0(P)$ is $4$-fold degenerate (see Lemma \ref{lm:Hph<H_0} in the appendix). The lowest eigenvalue of $H_r$ is given by 
\begin{align*}
e_0 = - \mu/2.
\end{align*}
The projection onto the vector space associated with $E_0(P)$ is denoted by
\begin{align*}
\Pi_{0,\ge\sigma}(P) := \mathds{1}_{ \{ E_0(P) \} }\big ( K_{0,\ge\sigma}(P) \big ), \quad \text{and} \quad \bar \Pi_{0,\ge\sigma}(P) := \mathds{1}- \Pi_{0,\ge\sigma}(P).
\end{align*}
Note that $\Pi_{0,\ge\sigma}(P)$ is independent of $P$. Setting $\Pi_{0,\ge\sigma} := \Pi_{0,\ge\sigma}(P)$, $\bar \Pi_{0,\ge\sigma} := \bar \Pi_{0,\ge\sigma}(P)$, we have
\begin{align*}
\Pi_{0,\ge\sigma} = \mathds{1} \otimes \pi_0 \otimes \Pi_{\Omega_{\ge\sigma}},
\end{align*}
where $\pi_0$ is the projection onto the vector space associated with the ground state $\phi_0$ of $H_r$, and $\Pi_{\Omega_{\ge\sigma}}$ is the projection onto the vacuum sector in $\mathcal{H}_{\ph,\ge\sigma}$.
Moreover, we set
\begin{equation*}
E_{g,\ge\sigma}(P) := E( K_{g,\ge\sigma}(P)).
\end{equation*}

As mention above, we will analyze the bottom of the spectrum of $K_{g, \ge \sigma} (P)$ iteratively, by letting the infrared cutoff parameter $\sigma \downarrow 0$. Of course, for $\sigma \ge \Lambda$, we have that $K_{g,\ge\sigma}(P) = K_{0,\ge\sigma}(P)$ and hence the spectrum of $K_{g,\ge\sigma}(P)$ is explicit: It is composed of the $4$-fold degenerate eigenvalue $E_{g,\ge\sigma}(P) = E_0(P)$ and a semi-axis of absolutely continuous spectrum $[E_{g,\ge\sigma}(P) + \mathrm{C} \sigma,\infty)$ for some positive constant $\mathrm{C}$ (see Figure 1).
\begin{figure}[htbp]
\label{fig:1}
\begin{center}
\setlength{\unitlength}{2000sp}%
\begingroup\makeatletter\ifx\SetFigFont\undefined%
\gdef\SetFigFont#1#2#3#4#5{%
  \reset@font\fontsize{#1}{#2pt}%
  \fontfamily{#3}\fontseries{#4}\fontshape{#5}%
  \selectfont}%
\fi\endgroup%
\begin{picture}(10312,1456)(1051,-5810)
\put(4306,-5661){\makebox(0,0)[lb]{\smash{{\SetFigFont{10}{14.4}{\rmdefault}{\mddefault}{\updefault}{\color[rgb]{0,0,0}$\mathrm{C} \sigma$}%
}}}}
\thinlines
{\color[rgb]{0,0,0}\multiput(1126,-4936)(6.00000,6.00000){26}{\makebox(1.6667,11.6667){\SetFigFont{5}{6}{\rmdefault}{\mddefault}{\updefault}.}}
}%
{\color[rgb]{0,0,0}\put(7701,-4636){\line( 0,-1){450}}
}%
{\color[rgb]{0,0,0}\put(7701,-5086){\line( 1, 0){ 75}}
}%
{\color[rgb]{0,0,0}\put(7701,-4636){\line( 1, 0){ 75}}
}%
{\color[rgb]{0,0,0}\put(7701,-4861){\line( 1, 0){3650}}
}%
{\color[rgb]{0,0,0}\put(1201,-5311){\vector(-1, 0){  0}}
\put(1201,-5311){\vector( 1, 0){6575}}
}%
\put(651,-4536){\makebox(0,0)[lb]{\smash{{\SetFigFont{10}{14.4}{\rmdefault}{\mddefault}{\updefault}{\color[rgb]{0,0,0}$E_{g,\ge\sigma}(P)$}%
}}}}
{\color[rgb]{0,0,0}\multiput(1276,-4936)(-6.00000,6.00000){26}{\makebox(1.6667,11.6667){\SetFigFont{5}{6}{\rmdefault}{\mddefault}{\updefault}.}}
}%
\end{picture}%
\caption{Spectrum of $K_{g,\ge\sigma}(P)$ for $\sigma \ge \Lambda$}
\end{center}
\end{figure}
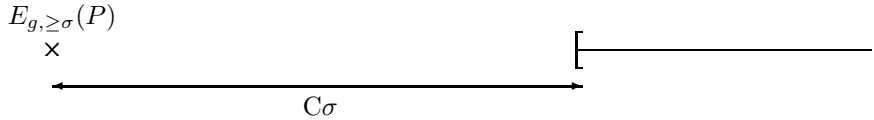

Compared to previous works, the main substantial difficulty we encounter comes from the fact that, as $\sigma$ becomes strictly less that $\Lambda$, $E_{g,\ge\sigma}(P)$ splits into $4$ (generally) distinct eigenvalues. Therefore, in particular, for $\sigma$ such that $g^2 \ll \sigma < \Lambda$, the gap above $E_{g,\ge\sigma}(P)$ in the spectrum of $K_{g,\ge\sigma}(P)$, $\mathrm{Gap}( K_{g,\ge\sigma}(P) )$, is negligible compared to $\sigma$. To overcome this difficulty, we have to start with analyzing the Hamiltonian $K_{g,\ge\sigma}(P)$ for $\sigma = \mathrm{C} g^2$ with $\mathrm{C}$ a suitably chosen positive constant. Since for these values of the parameters, the perturbation is of the same order as the distance between the ground state and the essential spectrum, one cannot straightforwardly apply usual perturbation theory. We then do second order perturbation theory with the help of the Feshbach-Schur map.

The Feshbach-Schur map is a natural tool to study second order perturbation of (possibly embedded) eigenvalues of self-adjoint operators. In the context of non-relativistic QED, it was introduced in \cite{BFS} and further developed in \cite{BCFS,GH}. The Feshbach-Schur operator we consider here is associated with $K_{g,\ge\sigma}(P) - E_{g,\ge\sigma}(P)$ and $\Pi_{0,\ge\sigma}$ and is defined by
\begin{align}
 F_{g,\ge\sigma}(P) :=
& ( E_0(P) - E_{g,\ge\sigma}(P) ) \Pi_{0,\ge\sigma}  - \Pi_{0,\ge\sigma} W_{g,\ge\sigma}(P) \phantom{ \big ]^{-1} }\notag \\
 &\big [ K_{0,\ge\sigma}(P) - E_{g,\ge\sigma}(P) + \bar \Pi_{0,\ge\sigma} W_{g,\ge\sigma}(P) \bar \Pi_{0,\ge\sigma} \big ]^{-1} \bar \Pi_{0,\ge\sigma} W_{g,\ge\sigma}(P) \Pi_{0,\ge\sigma}. \label{eq:FSoperator}
\end{align} 
We shall see in Section \ref{section:2} that this operator is well-defined for suitable values of the parameters. The main property of $F_{g,\ge\sigma}(P)$ that we shall use is that 
\begin{align*}
K_{g,\ge\sigma}(P) - E_{g,\ge\sigma}(P) \ge F_{g,\ge\sigma}(P),
\end{align*}
(see Lemma \ref{lm:term_order_g^2}). Combined with the min-max principle, this operator inequality appears to be very useful in the context of the present paper. In particular, it will allow us to prove that the spectrum of $K_{g,\ge\sigma}(P)$ has the form pictured in Figure 2 (see Section \ref{section:2}): The bottom of the spectrum, $E_{g,\ge\sigma}(P)$, is an isolated eigenvalue separated by a gap of size $\delta g^2$ from three other eigenvalues and a semi-axis of essential spectrum.
\begin{figure}[htbp]
\begin{center}
\setlength{\unitlength}{2000sp}%
\begingroup\makeatletter\ifx\SetFigFont\undefined%
\gdef\SetFigFont#1#2#3#4#5{%
  \reset@font\fontsize{#1}{#2pt}%
  \fontfamily{#3}\fontseries{#4}\fontshape{#5}%
  \selectfont}%
\fi\endgroup%
\begin{picture}(10312,1456)(1051,-5810)
\put(1900,-5661){\makebox(0,0)[lb]{\smash{{\SetFigFont{10}{14.4}{\rmdefault}{\mddefault}{\updefault}{\color[rgb]{0,0,0}$\delta g^2$}%
}}}}
\thinlines
{\color[rgb]{0,0,0}\multiput(1126,-4936)(6.00000,6.00000){26}{\makebox(1.6667,11.6667){\SetFigFont{5}{6}{\rmdefault}{\mddefault}{\updefault}.}}
}%
{\color[rgb]{0,0,0}\multiput(3126,-4936)(6.00000,6.00000){26}{\makebox(1.6667,11.6667){\SetFigFont{5}{6}{\rmdefault}{\mddefault}{\updefault}.}}
}%
{\color[rgb]{0,0,0}\multiput(3626,-4936)(6.00000,6.00000){26}{\makebox(1.6667,11.6667){\SetFigFont{5}{6}{\rmdefault}{\mddefault}{\updefault}.}}
}%
{\color[rgb]{0,0,0}\multiput(4126,-4936)(6.00000,6.00000){26}{\makebox(1.6667,11.6667){\SetFigFont{5}{6}{\rmdefault}{\mddefault}{\updefault}.}}
}%
{\color[rgb]{0,0,0}\put(5001,-4636){\line( 0,-1){450}}
}%
{\color[rgb]{0,0,0}\put(5001,-5086){\line( 1, 0){ 75}}
}%
{\color[rgb]{0,0,0}\put(5001,-4636){\line( 1, 0){ 75}}
}%
{\color[rgb]{0,0,0}\put(5001,-4861){\line( 1, 0){6350}}
}%
{\color[rgb]{0,0,0}\put(1201,-5311){\vector(-1, 0){  0}}
\put(1201,-5311){\vector( 1, 0){2000}}
}%
\put(651,-4536){\makebox(0,0)[lb]{\smash{{\SetFigFont{10}{14.4}{\rmdefault}{\mddefault}{\updefault}{\color[rgb]{0,0,0}$E_{g,\ge\sigma}(P)$}%
}}}}
{\color[rgb]{0,0,0}\multiput(1276,-4936)(-6.00000,6.00000){26}{\makebox(1.6667,11.6667){\SetFigFont{5}{6}{\rmdefault}{\mddefault}{\updefault}.}}
}%
{\color[rgb]{0,0,0}\multiput(3276,-4936)(-6.00000,6.00000){26}{\makebox(1.6667,11.6667){\SetFigFont{5}{6}{\rmdefault}{\mddefault}{\updefault}.}}
}%
{\color[rgb]{0,0,0}\multiput(3776,-4936)(-6.00000,6.00000){26}{\makebox(1.6667,11.6667){\SetFigFont{5}{6}{\rmdefault}{\mddefault}{\updefault}.}}
}%
{\color[rgb]{0,0,0}\multiput(4276,-4936)(-6.00000,6.00000){26}{\makebox(1.6667,11.6667){\SetFigFont{5}{6}{\rmdefault}{\mddefault}{\updefault}.}}
}%
\end{picture}%
\caption{Spectrum of $K_{g,\ge\sigma}(P)$ for $\sigma = \mathrm{C} g^2$, $\mathrm{C} \gg 1$}
\end{center}
\end{figure}
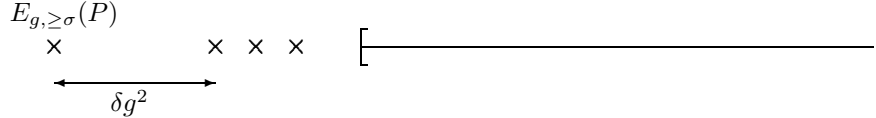

The rest of the proof borrows ideas from \cite{Pizzo,BFP,AFGG2}. Namely, we shall prove (in Section \ref{section:3}) that if the ground state of $K_{g,\ge\sigma}(P)$ is unique and if $\mathrm{Gap}( K_{g,\ge\sigma}(P) ) \ge \eta \sigma$, then the same holds for $K_{g,\ge\tau}(P)$ with $\tau = \kappa \sigma$, $0 < \kappa < 1$. This will show that for small $\sigma$'s, the spectrum of $K_{g,\ge\sigma}(P)$ has the form pictured in Figure 3 (a non-degenerate eigenvalue separated from a gap of size $\eta \sigma$ from the rest of the spectrum). At the end, the parameters $\eta$ and $\kappa$ will have to be carefully chosen, in relation with the initial analysis of Section \ref{section:2}.
\begin{figure}[htbp]
\begin{center}
\setlength{\unitlength}{2000sp}%
\begingroup\makeatletter\ifx\SetFigFont\undefined%
\gdef\SetFigFont#1#2#3#4#5{%
  \reset@font\fontsize{#1}{#2pt}%
  \fontfamily{#3}\fontseries{#4}\fontshape{#5}%
  \selectfont}%
\fi\endgroup%
\begin{picture}(10312,1456)(1051,-5810)
\put(1710,-5661){\makebox(0,0)[lb]{\smash{{\SetFigFont{10}{14.4}{\rmdefault}{\mddefault}{\updefault}{\color[rgb]{0,0,0}$\eta\sigma$}%
}}}}
\thinlines
{\color[rgb]{0,0,0}\multiput(1126,-4936)(6.00000,6.00000){26}{\makebox(1.6667,11.6667){\SetFigFont{5}{6}{\rmdefault}{\mddefault}{\updefault}.}}
}%
{\color[rgb]{0,0,0}\put(2501,-4636){\line( 0,-1){450}}
}%
{\color[rgb]{0,0,0}\put(2501,-5086){\line( 1, 0){ 75}}
}%
{\color[rgb]{0,0,0}\put(2501,-4636){\line( 1, 0){ 75}}
}%
{\color[rgb]{0,0,0}\put(2501,-4861){\line( 1, 0){8850}}
}%
{\color[rgb]{0,0,0}\put(1201,-5311){\vector(-1, 0){  0}}
\put(1201,-5311){\vector( 1, 0){1300}}
}%
\put(651,-4536){\makebox(0,0)[lb]{\smash{{\SetFigFont{10}{14.4}{\rmdefault}{\mddefault}{\updefault}{\color[rgb]{0,0,0}$E_{g,\ge\sigma}(P)$}%
}}}}
{\color[rgb]{0,0,0}\multiput(1276,-4936)(-6.00000,6.00000){26}{\makebox(1.6667,11.6667){\SetFigFont{5}{6}{\rmdefault}{\mddefault}{\updefault}.}}
}%
\end{picture}%
\caption{Spectrum of $K_{g,\ge\sigma}(P)$ for $\sigma \le \mathrm{C}' g^2$, $\mathrm{C}' \ll 1$}
\end{center}
\end{figure}

We conclude this section with introducing a few more notations related to the infrared decomposition, which will be useful in Section \ref{section:3}. 
For $0 \le \tau \le \sigma$, let
\begin{align*}
& \mathcal{H}_{\ph,\ge \tau}^{\le\sigma} := \mathbb{C} \oplus \bigoplus_{n=1}^\infty S_n \left [ \mathrm{L}^2( \{ (k,\lambda) \in \mathbb{R}^3 \times \{1,2\} , \tau \le |k| \le \sigma \} )^{ \otimes^n} \right ].
\end{align*}
The vacuum in $\mathcal{H}_{\ph,\ge \tau}^{\le\sigma}$ is denoted by $\Omega_{\ge\tau}^{\le\sigma}$ and the projection onto the vacuum sector is denoted by $\Pi_{ \Omega_{\ge\tau}^{\le\sigma} }$. The Hilbert spaces $\mathcal{H}_{\fib,\ge\tau}$ and $\mathcal{H}_{\fib,\ge\sigma} \otimes \mathcal{H}_{\ph,\ge\tau}^{\le\sigma}$ are isomorphic. We shall sometimes not distinguish between the two of them. As an operator on $\mathcal{H}_{\fib,\ge\tau}$, we set
\begin{align}
W_{g,\ge\tau}^{\le\sigma}(P) =& - \frac{ g }{ m_\el } \text{:} \left ( \big ( \frac{ m_\el }{ M } ( P - P_{\ph,\ge\tau} ) + p_r - g A_{\ge\sigma} ( \frac{ m_\el }{ M } g^{\frac{2}{3}} r ) \big ) \cdot A_{\ge\tau}^{\le\sigma} ( \frac{ m_\el }{ M } g^{\frac{2}{3}} r ) \right ) \text{:} \notag \\ 
& + \frac{ g }{ m_\n } \text{:} \left ( \big ( \frac{ m_\n }{ M } ( P - P_{\ph,\ge\tau} ) - p_r + g A_{\ge\sigma} ( - \frac{ m_\n }{ M } g^{\frac{2}{3}} r ) \big ) \cdot A_{\ge\tau}^{\le\sigma} ( - \frac{ m_\n }{ M } g^{\frac{2}{3}} r ) \right ) \text{:} \notag \\ 
& + \frac{ g^2 }{ 2m_\el } \text{:} A_{\ge\tau}^{\le\sigma} ( \frac{ m_\el }{ M } g^{\frac{2}{3}} r )^2 \text{:} + \frac{ g^2 }{ 2m_\n } \text{:} A_{\ge\tau}^{\le\sigma} ( - \frac{ m_\n }{ M } g^{\frac{2}{3}} r )^2 \text{:} \notag \\
& -\frac{g}{2m_\el} \sigma^{\el} \cdot B_{\ge\tau}^{\le\sigma}( \frac{ m_\el }{ M } g^{\frac{2}{3}} r ) + \frac{g}{2m_\n} \sigma^{\n} \cdot B_{\ge\tau}^{\le\sigma} ( - \frac{ m_\n }{ M } g^{\frac{2}{3}} r ), \label{eq:Wg>tau<sigma}
\end{align}
where $A_{\ge\tau}^{\le\sigma}(\cdot)$ and $B_{\ge\tau}^{\le\sigma}(\cdot)$ are given by the same expressions as $A(\cdot)$ and $B(\cdot)$ respectively, except that the integrals are taken over $\{ k \in \mathbb{R}^3 , \tau \le |k| \le \sigma \}$. Note that
\begin{align*}
H_{g,\ge\tau}(P) = H_{g,\ge\sigma}(P) + W_{g,\ge\tau}^{\le\sigma}(P).
\end{align*}
Finally, in the case where $\tau=0$, the subindex $\ge 0$ is removed from the notations above, that is, for instance, $H_{\ph}^{\le\sigma} := H_{\ph, \ge 0}^{\le\sigma}$, $W_g^{\le\sigma}(P) := W_{g,\ge0}^{\le\sigma}(P)$ and so on.

Throughout the paper, the notation $\lesssim \cdots$ will stand for $\le \mathrm{C} \cdots$ where $\mathrm{C}$ is a positive constant independent of the parameters. For any vector $v$, $[v]$ and $[v]^\perp$ will denote respectively the subspace spanned by $v$ and its orthogonal complement.

%%%%%%%%%%%%%%%%%%%%%%%%%%%%%%%%%%%%
%%%%%%%%%%%%%%%%%%%%%%%%%%%%%%%%%%%%
%%%%%%%%%%%%%%%%%%%%%%%%%%%%%%%%%%%%
%%%%%%%%%%%%%%%%%%%%%%%%%%%%%%%%%%%%
\section{Existence of a gap for large enough infrared cutoffs}\label{section:2}

In this section, we investigate the spectrum of the infrared cutoff Hamiltonian $K_{g,\ge\sigma}(P)$ for values of the coupling constant $g$ and of the infrared cutoff parameter $\sigma$ fixed such that $\beta_{c2}^{-1} g^2 \le \sigma \le \beta_{c1}^{-1} g^2$ (for some $0 < \beta_{c1} < \beta_{c2}$ to be determined later).

For any small enough $g$ and $P$, and for any $\sigma,\eta > 0$, let $\mathbf{Gap}(g,P,\sigma,\eta)$ denote the following assertion:
$$
\mathbf{Gap}(g,P,\sigma,\eta)
\left \{ 
\begin{array}{cl}
 &\mathrm{(i)} \phantom{i} \quad E_{g,\ge\sigma}(P) \text{ is a simple eigenvalue of } K_{g,\ge\sigma}(P), \vspace{0,1cm} \\
 &\mathrm{(ii)} \quad \mathrm{Gap}( K_{g,\ge\sigma}(P) ) \ge \eta \sigma.
\end{array}
\right.
$$

The main result of this section is the following.

\begin{theorem}\label{thm:large_sigma}
There exist $p_c>0$, $\beta_{c2}>0$ and $\delta > 0$ such that, for all $0 < \beta_{c1} < \beta_{c2}$, there exists $g_c > 0$ such that, for all $0 \le |P| \le p_c$, $0 < g \le g_c$ and $\beta_{c1} \le \beta \le \beta_{c2}$,
\begin{align*}
\mathbf{Gap}( g , P , \sigma , \delta \beta ) \text{ holds},
\end{align*}
where $\sigma = g^2 \beta^{-1}$.
\end{theorem}
The statement of Theorem \ref{thm:large_sigma} expresses the fact that, for small enough values of the coupling constant $g$ and total momentum $P$, there is a gap at least of size $\delta g^2$ in the spectrum of $K_{g,\ge\sigma}(P)$ above the non-degenerate ground state eigenvalue $E_{g,\ge\sigma}(P)$, provided that the infrared cutoff parameter $\sigma$ obeys $\beta_{c2}^{-1} g^2 \le \sigma \le \beta_{c1}^{-1} g^2$.

\subsection{Preliminary lemmas}
The proof of Theorem \ref{thm:large_sigma} relies on a few lemmas that we shall establish in this preliminary subsection. For the convenience of the reader, some standard estimates used several times in the proofs below are recalled in the appendix.

We begin with verifying that the Feshbach-Schur operator $F_{g,\ge\sigma}(P)$ introduced in \eqref{eq:FSoperator} is well-defined for suitable values of the parameters.
\begin{lemma}\label{lm:term_order_g^2}
There exist $p_c>0$ and $\beta_{c2} > 0$ such that, for all $0 < \beta_{c1} < \beta_{c2}$, there exists $g_c > 0$ such that, for all $0 < g \le g_c$, $0 \le |P| \le p_c$ and $\beta_{c1} \le \beta \le \beta_{c2}$, the Feshbach-Schur operator defined in \eqref{eq:FSoperator}, $F_{g,\ge\sigma}(P)$ (where $\sigma = g^2 \beta^{-1}$), is a bounded operator on $\mathrm{Ran}( \Pi_{0,\ge\sigma}) \subset \mathcal{H}_{\mathrm{fib}}$ given by
\begin{align}
&F_{g,\ge\sigma}(P) = \big ( E_0(P) - E_{g,\ge\sigma}(P) ) \big ) \Pi_{0,\ge\sigma}  \notag \\
& \quad - \Pi_{0,\ge\sigma} W_{g,\ge\sigma}(P) \big [ K_{0,\ge\sigma}(P) - E_{g,\ge\sigma}(P) \big ]^{-1} \bar \Pi_{0,\ge\sigma} W_{g,\ge\sigma}(P) \Pi_{0,\ge\sigma} \notag \\
& \quad + \mathrm{Rem}_1(g,P,\beta), \phantom{ \big ]^{-1} } \label{eq:a2}
\end{align}
where $\mathrm{Rem}_1(g,P,\beta)$ is a bounded operator on $\mathrm{Ran}( \Pi_{0,\ge\sigma})$ satisfying
\begin{align}\label{eq:a1}
\| \mathrm{Rem}_1(g,P,\beta) \| \lesssim g^2 \beta^{\frac{1}{2}}.
\end{align}
Moreover the following inequality holds in the sense of quadratic forms on $D( K_{g,\ge\sigma}(P) )$:
\begin{align}\label{eq:ineq_Fesh}
K_{g,\ge\sigma}(P) - E_{g,\ge\sigma}(P) \ge F_{g,\ge\sigma}(P).
\end{align}
\end{lemma}
\begin{proof}
Let $g_c$, $p_c$, $\sigma_c$ and $\mathrm{C}_W$ be given by Lemma \ref{lm:estimate_Wg}. Let $\beta_{c2}$ be such that $\beta_{c2}^{1/2} \le (6 \mathrm{C}_W)^{-1}$, and let $0 < \beta_{c1} < \beta_{c2}$. Possibly by considering a smaller $g_c$, we can assume that $g_c^2 \le \beta_{c1} \sigma_c$ and hence, for all $0 < g \le g_c$ and $\beta \ge \beta_{c1}$, we have that $0 < \sigma = g^2\beta^{-1} \le \sigma_c$. In addition we impose that $p_c \le M / 2$.

Fix $g$, $P$ and $\beta$ as in the statement of the lemma. By Lemmas \ref{lm:Hph<H_0} and \ref{lm:Eg-E0}, we have that $K_{0,\ge\sigma}(P) - E_{g,\ge\sigma}(P)$ is bounded invertible on $\mathrm{Ran}( \bar \Pi_{0,\ge\sigma} )$ and satisfies
\begin{align*}
\big ( K_{0,\ge\sigma}(P) - E_{g,\ge\sigma}(P)  \big ) \bar \Pi_{0,\ge\sigma} & \ge \big ( E_0(P) - E_{g,\ge\sigma}(P) + ( 1 - \frac{ |P| }{ M } ) \sigma \big ) \bar \Pi_{0,\ge\sigma} \ge \frac{ \sigma }{2} \bar \Pi_{0,\ge\sigma}.
\end{align*}
Using again that $E_{g,\ge\sigma}(P) \le E_0(P)$ by Lemma \ref{lm:Eg-E0}, it then follows from a straightforward application of the Spectral Theorem that
\begin{align}
&\Big \| \big [ K_{0,\ge\sigma}(P) - E_{g,\ge\sigma}(P)  \big ]^{-1} \bar \Pi_{0,\ge\sigma} \big ( K_{0,\ge\sigma}(P) - E_0(P) + \sigma \big ) \Big \| \notag \\
& \le \Big \| \big [ K_{0,\ge\sigma}(P) - E_{g,\ge\sigma}(P)  \big ]^{-1} \bar \Pi_{0,\ge\sigma} \big ( K_{0,\ge\sigma}(P) - E_{g,\ge\sigma}(P) + \sigma \big ) \Big \| \le 3. \label{eq:Neumann_estimate_1}
\end{align}
Moreover, by Lemma \ref{lm:estimate_Wg}, we have that
\begin{align}
& \Big \| \big [ K_{0,\ge\sigma}(P) - E_0(P) + \sigma \big ]^{-\frac{1}{2}} W_{g,\ge\sigma}(P) \big [ K_{0,\ge\sigma}(P) - E_0(P) + \sigma \big ]^{-\frac{1}{2}} \Big \| \notag \\
& \le \mathrm{C}_W g \sigma^{-\frac{1}{2}} = \mathrm{C}_W \beta^{\frac{1}{2}}, \label{eq:Neumann_estimate_2}
\end{align}
and hence
\begin{align*}
& \Big \| \big [ K_{0,\ge\sigma}(P) - E_{g,\ge\sigma}(P)  \big ]^{-\frac{1}{2}} \bar \Pi_{0,\ge\sigma} W_{g,\ge\sigma}(P) \big [ K_{0,\ge\sigma}(P) - E_{g,\ge\sigma}(P)  \big ]^{-\frac{1}{2}} \bar \Pi_{0,\ge\sigma} \Big \| \le 3 \mathrm{C}_W \beta^{\frac{1}{2}}. 
\end{align*} 
Taking into account the choice of $\beta_{c2}$, we have $ 3 \mathrm{C}_W \beta^{\frac{1}{2}} \le 1/2$ so that in particular
\begin{align}
& \left ( K_{0,\ge\sigma}(P) - E_{g,\ge\sigma}(P)  \right ) \bar \Pi_{0,\ge\sigma} + \bar \Pi_{0,\ge\sigma} W_{g,\ge\sigma}(P) \bar \Pi_{0,\ge\sigma} \notag \\
& \ge \frac{1}{2} \left ( K_{0,\ge\sigma}(P) - E_{g,\ge\sigma}(P)  \right ) \bar \Pi_{0,\ge\sigma} \ge \frac{\sigma}{4}\bar \Pi_{0,\ge\sigma}. \label{eq:H>F}
\end{align}
Therefore the operator $( K_{0,\ge\sigma}(P) - E_{g,\ge\sigma}(P)  ) \bar \Pi_{0,\ge\sigma} + \bar \Pi_{0,\ge\sigma} W_{g,\ge\sigma}(P) \bar \Pi_{0,\ge\sigma}$ is bounded invertible on $\mathrm{Ran} ( \bar \Pi_{0,\ge\sigma})$. Using in addition that $W_{g,\ge\sigma}(P)$ is relatively bounded with respect to $K_{0,\ge\sigma}(P)$, we obtain that $F_{g,\ge\sigma}(P)$ is indeed a well-defined bounded operator on $\mathrm{Ran}( \Pi_{0,\ge\sigma})$.

Next, using again \eqref{eq:Neumann_estimate_1} and \eqref{eq:Neumann_estimate_2}, we obtain
\begin{align}
\Big \| \Pi_{0,\ge\sigma} W_{g,\ge\sigma}(P) \big [ K_{0,\ge\sigma}(P) - E_{g,\ge\sigma}(P)  \big ]^{-\frac{1}{2}} \bar \Pi_{0,\ge\sigma} \Big \| & \le 2 \mathrm{C}_W \beta^{\frac{1}{2}} \sigma^{\frac{1}{2}} = 2 \mathrm{C}_W g. \label{eq:Neumann_estimate_3}
\end{align}
A standard Neumann series decomposition together with the previous estimates then lead to
\begin{align*}
& \Pi_{0,\ge\sigma} W_{g,\ge\sigma}(P) \big [ K_{0,\ge\sigma}(P) - E_{g,\ge\sigma}(P)  + \bar \Pi_{0,\ge\sigma} W_{g,\ge\sigma}(P) \bar \Pi_{0,\ge\sigma} \big ]^{-1} \bar \Pi_{0,\ge\sigma} W_{g,\ge\sigma}(P) \Pi_{0,\ge\sigma} \phantom{ \big ]^{-1} } \notag \\
& = \Pi_{0,\ge\sigma} W_{g,\ge\sigma}(P) \big [ K_{0,\ge\sigma}(P) - E_{g,\ge\sigma}(P)  \big ]^{-1} \bar \Pi_{0,\ge\sigma} W_{g,\ge\sigma}(P) \Pi_{0,\ge\sigma} + \mathrm{Rem}_1(g,P,\beta) , \phantom{ \big ]^{-1} }
\end{align*}
where $\mathrm{Rem}_1(g,P,\beta)$ is a bounded operator on $\mathrm{Ran}( \Pi_{0,\ge\sigma})$ satisfying \eqref{eq:a1}. 

Finally, to prove \eqref{eq:ineq_Fesh}, it suffices to use the following identity
\begin{align}\label{eq:H>F2}
&K_{g,\ge\sigma}(P) - E_{g,\ge\sigma}(P) = F_{g,\ge\sigma}(P) + R^* R,
\end{align}
where
\begin{align*}
R :=  \big [ K_{0,\ge\sigma}(P) - E_{g,\ge\sigma}(P)  + \bar \Pi_{0,\ge\sigma} W_{g,\ge\sigma}(P) \bar \Pi_{0,\ge\sigma} \big ]^{-\frac{1}{2}} \bar \Pi_{0,\ge\sigma} ( K_{g,\ge\sigma}(P) - E_{g,\ge\sigma}(P) ).
\end{align*}
We observe that the operator square root appearing in the expression of $R$ is well-defined by \eqref{eq:H>F}. Equation \eqref{eq:H>F2} follows from straightforward algebraic computations (see e.g. \cite{BCFS,GH}). This concludes the proof of the lemma.
\end{proof}
Our next task is to extract the second order term from \eqref{eq:a2}. It is the purpose of the following three lemmas.
\begin{lemma}\label{lm:term_order_g^2_lm1}
There exist $g_c>0$, $p_c>0$ and $\sigma_c > 0$ such that, for all $0 \le g \le g_c$, $0 \le |P| \le p_c$ and $0 < \sigma \le \sigma_c$,
\begin{align}
& \Pi_{0,\ge\sigma} W_{g,\ge\sigma}(P) \big [ K_{0,\ge\sigma}(P) - E_{g,\ge\sigma}(P) \big ]^{-1} \bar \Pi_{0,\ge\sigma} W_{g,\ge\sigma}(P) \Pi_{0,\ge\sigma} \notag \\
&= \sum_{\lambda=1,2} \int_{\mathbb{R}^3} \Pi_{0,\ge\sigma} \tilde w_{\ge\sigma}(r,k,\lambda) \big [ H_r + \frac{1}{2M} ( P - k )^2 + |k| - E_{g,\ge\sigma}(P) \big ]^{-1} w_{\ge\sigma}(r,k,\lambda) \Pi_{0,\ge\sigma} \d k \notag \\
&\quad + \mathrm{Rem}_2(g,P,\sigma) , \label{eq:F(epsilon)_3}
\end{align}
where 
\begin{align}
w_{\ge\sigma}(r,k,\lambda) := & - \frac{ g }{ m_\el } \left ( \big ( \frac{ m_\el }{ M } ( P - P_{\ph,\ge\sigma} ) + p_r \big ) \cdot h^A_{\ge\sigma} ( \frac{ m_\el }{ M } g^{\frac{2}{3}} r , k , \lambda ) \right ) \notag \\
& + \frac{ g }{ m_\n } \left ( \big ( \frac{ m_\n }{ M } ( P - P_{\ph,\ge\sigma} ) - p_r \big ) \cdot h^A_{\ge\sigma} ( - \frac{ m_\n }{ M } g^{\frac{2}{3}} r , k , \lambda ) \right ) \notag \\
& - \frac{g}{2m_\el} \sigma^{\el} \cdot h^B_{\ge\sigma}( \frac{ m_\el }{ M } g^{\frac{2}{3}} r , k , \lambda ) + \frac{g}{2m_\n} \sigma^{\n} \cdot h^B_{\ge\sigma} ( - \frac{ m_\n }{ M } g^{\frac{2}{3}} r , k , \lambda ), \label{eq:w(r,k,lambda)}
\end{align}
$\tilde w_{\ge\sigma}(r,k,\lambda)$ is given by the same expression except that $h^A_{\ge\sigma}$, $h^B_{\ge\sigma}$ are replaced by their conjugate $\bar{h}^A_{\ge\sigma}$, $\bar{h}^B_{\ge\sigma}$, 
and $\mathrm{Rem}_2(g,P,\sigma)$ is a bounded operator on $\mathrm{Ran}( \Pi_{0,\ge\sigma})$ satisfying
\begin{align*}
\| \mathrm{Rem}_2(g,P,\sigma) \| \lesssim g^3.
\end{align*}
\end{lemma}
\begin{proof}
It suffices to introduce the expression \eqref{eq:Wg>sigma} of $W_{g,\ge\sigma}(P)$ into the operator
\begin{align*}
\Pi_{0,\ge\sigma} W_{g,\ge\sigma}(P) \big [ K_{0,\ge\sigma}(P) - E_{g,\ge\sigma}(P) \big ]^{-1} \bar \Pi_{0,\ge\sigma} W_{g,\ge\sigma}(P) \Pi_{0,\ge\sigma},
\end{align*}
and next to estimate each term separately. An explicit computation then leads directly to the statement of the lemma (see the proof of Lemma A.9 in \cite{AF2} for more details).
\end{proof}
\begin{lemma}\label{lm:term_order_g^2_lm2}
There exist $g_c>0$, $p_c>0$ and $\sigma_c > 0$ such that, for all $0 \le g \le g_c$, $0 \le |P| \le p_c$, $0 < \sigma \le \sigma_c$ and $\lambda \in \{1,2\}$,
\begin{align*}
&\int_{\mathbb{R}^3} \Pi_{0,\ge\sigma} \tilde w_{\ge\sigma}(r,k,\lambda) \big [ H_r + \frac{1}{2M} ( P - k )^2 + |k| - E_{g,\ge\sigma}(P) \big ]^{-1} w_{\ge\sigma}(r,k,\lambda) \Pi_{0,\ge\sigma} \d k \notag \\
&= \int_{\mathbb{R}^3} \Pi_{0,\ge\sigma} \tilde w_{\ge\sigma}(0,k,\lambda) \big [ H_r + \frac{1}{2M} ( P - k )^2 + |k| - E_{g,\ge\sigma}(P) \big ]^{-1} w_{\ge\sigma}(0,k,\lambda) \Pi_{0,\ge\sigma} \d k \notag \\
&\quad + \mathrm{Rem}_3(g,P,\sigma,\lambda) , \phantom{\int} 
\end{align*}
where $w_{\ge\sigma}(0,k,\lambda)$ and $\tilde w_{\ge\sigma}(0,k,\lambda)$ are defined by \eqref{eq:w(r,k,lambda)}, and $\mathrm{Rem}_3(g,P,\sigma,\lambda)$ is a bounded operator on $\mathrm{Ran}( \Pi_{0,\ge\sigma})$ satisfying
\begin{align*}
\| \mathrm{Rem}_3(g,P,\sigma,\lambda) \| \lesssim g^{\frac{8}{3}}.
\end{align*}
\end{lemma}
\begin{proof}
It follows from the definitions \eqref{eq:h^A}--\eqref{eq:h^B} of $h_j^A$ and $h^B_j$ that
\begin{align*}
&\big | h^A_{j,\ge\sigma}( r, k ,\lambda ) - h^A_{j,\ge\sigma}( 0 , k, \lambda ) \big | \lesssim \mathds{1}_{ |k| \ge \sigma }(k) |k|^{\frac{1}{2}} \chi_\Lambda(k) |r|, \\
&\big | h^B_{j,\ge\sigma}( r, k ,\lambda ) - h^B_{j,\ge\sigma}( 0 , k, \lambda ) \big | \lesssim \mathds{1}_{ |k| \ge \sigma }(k) |k|^{\frac{3}{2}} \chi_\Lambda(k) |r|,
\end{align*}
for any $j \in \{1,2,3\}$, $\lambda \in \{1,2\}$, $r \in \mathbb{R}^3$ and $k \in \mathbb{R}^3$. This implies that
\begin{align*}
& \Big \|\Pi_{0,\ge\sigma} \big ( \frac{ m_\el }{ M } ( P - P_{\ph,\ge\sigma} ) + p_r \big ) \cdot \big ( h^A_{\ge\sigma} ( \frac{ m_\el }{ M } g^{\frac{2}{3}} r , k , \lambda ) - h^A_{\ge\sigma} (0 , k , \lambda ) \big ) \Big \| \notag \\
& \lesssim g^{\frac{2}{3}} \mathds{1}_{ |k| \ge \sigma }(k) |k|^{\frac{1}{2}} \chi_\Lambda(k) \Big \| \Pi_{0,\ge\sigma} \big ( \frac{ m_\el }{ M } ( P - P_{\ph,\ge\sigma} ) + p_r \big ) \langle r \rangle \Big \| \notag \\
& \lesssim g^{\frac{2}{3}} \mathds{1}_{ |k| \ge \sigma }(k) |k|^{\frac{1}{2}} \chi_\Lambda(k).
\end{align*}
In the last inequality, we used in particular that $\| \langle r \rangle p_r \pi_0 \| < \infty$, where, recall, $\pi_0$ is the projection onto the ground state of the Schr\"odinger operator $H_r$. Similarly,
\begin{align*}
& \Big \|\Pi_{0,\ge\sigma} \sigma^\el \cdot \big ( h^B_{\ge\sigma} ( \frac{ m_\el }{ M } g^{\frac{2}{3}} r , k , \lambda ) - h^B_{\ge\sigma} (0 , k , \lambda ) \big ) \Big \| \notag \\
& \lesssim g^{\frac{2}{3}} \mathds{1}_{ |k| \ge \sigma }(k) |k|^{\frac{3}{2}} \chi_\Lambda(k) \Big \| \Pi_{0,\ge\sigma} \sigma^\el \langle r \rangle \Big \| \lesssim g^{\frac{2}{3}} \mathds{1}_{ |k| \ge \sigma }(k) |k|^{\frac{3}{2}} \chi_\Lambda(k).
\end{align*}
The same holds if $m_\el$ is replaced by $-m_\n$ and $\sigma^\el$ is replaced by $\sigma^\n$. Besides, using that $H_r + P^2 / 2M \ge E_0(P) \ge E_{g,\ge\sigma}(P)$ by Lemma \ref{lm:Eg-E0}, we obtain that
\begin{align*}
& \Big \| \big [ H_r + \frac{1}{2M} ( P - k )^2 + |k| - E_{g,\ge\sigma}(P) \big ]^{-1} \mathds{1}_{|k|\ge\sigma}(k) \Big \| \notag \\
&\le \frac{1}{ - ( k \cdot P ) / M + k^2 / 2M + |k| } \mathds{1}_{|k|\ge\sigma}(k) \le \frac{2}{ |k| } \mathds{1}_{|k|\ge\sigma}(k),
\end{align*}
for $|P| \le M / 2$, and hence the statement of the lemma easily follows.
\end{proof}
\begin{lemma}\label{lm:large5}
There exist $g_c>0$, $p_c>0$ and $\sigma_c > 0$ such that, for all $0 \le g \le g_c$, $0 \le |P| \le p_c$, $0 < \sigma \le \sigma_c$, and $\lambda \in \{1,2\}$,
\begin{align*}
&\int_{\mathbb{R}^3} \Pi_{0,\ge\sigma} \tilde w_{\ge\sigma}(0,k,\lambda) \big [ H_r + \frac{1}{2M} ( P - k )^2 + |k| - E_{g,\ge\sigma}(P) \big ]^{-1} w_{\ge\sigma}(0,k,\lambda) \Pi_{0,\ge\sigma} \d k \notag \\
&= g^2 \big ( \Gamma^{A,\mathrm{diag}}_{\ge\sigma}(P,\lambda) + \Gamma^B_{\ge\sigma}(P,\lambda) \big ) +  \mathrm{Rem}_4(g,P,\sigma,\lambda) , \phantom{\int}
\end{align*}
where
\begin{align*}
\Gamma^{A,\mathrm{diag}}_{\ge\sigma}(P,\lambda) :=& \frac{1}{\mu^2} \int_{\mathbb{R}^3} \Pi_{0,\ge\sigma} p_r \cdot h^A_{\ge\sigma} ( 0 , k , \lambda ) \big [ H_r + \frac{1}{2M} ( P - k )^2 + |k| - E_0(P) \big ]^{-1} \notag \\
& \phantom{ \frac{1}{\mu^2} \int_{\mathbb{R}^3} } ( \mathds{1} \otimes \bar \pi_0 \otimes \mathds{1} ) p_r \cdot h^A_{\ge\sigma} ( 0 , k , \lambda ) \Pi_{0,\ge\sigma} \d k,
\end{align*}
and
\begin{align*}
\Gamma^B_{\ge\sigma}(P,\lambda) :=& \int_{\mathbb{R}^3} \Pi_{0,\ge\sigma} \Big ( - \frac{1}{2m_\el} \sigma^{\el} \cdot \bar h^B_{\ge\sigma}( 0 , k , \lambda ) + \frac{1}{2m_\n} \sigma^{\n} \cdot \bar h^B_{\ge\sigma} ( 0 , k , \lambda ) \Big ) \notag \\
&\phantom{ \quad  \int_{\mathbb{R}^3} } \big [ e_0 + \frac{1}{2M} ( P - k )^2 + |k| - E_0(P) \big ]^{-1} \notag \\
& \phantom{ \quad \int_{\mathbb{R}^3} } \Big ( - \frac{1}{2m_\el} \sigma^{\el} \cdot h^B_{\ge\sigma}( 0 , k , \lambda ) + \frac{1}{2m_\n} \sigma^{\n} \cdot h^B_{\ge\sigma} ( 0 , k , \lambda ) \Big ) \Pi_{0,\ge\sigma}\d k. %\label{eq:GammaBsigma}
\end{align*}
Moreover $\mathrm{Rem}_4(g,P,\sigma,\lambda)$ is a bounded operator on $\mathrm{Ran}( \Pi_{0,\ge\sigma})$ satisfying
\begin{align*}
\| \mathrm{Rem}_4(g,P,\sigma,\lambda) \| \lesssim g^4.
\end{align*}
\end{lemma}
\begin{proof}
It follows from \eqref{eq:w(r,k,lambda)} that
\begin{align*}
w_{\ge \sigma}( 0 , k , \lambda) = & - \frac{ g }{ \mu } p_r \cdot h^A_{\ge\sigma} ( 0 , k , \lambda ) - \frac{g}{2m_\el} \sigma^{\el} \cdot h^B_{\ge\sigma}( 0 , k , \lambda ) + \frac{g}{2m_\n} \sigma^{\n} \cdot h^B_{\ge\sigma} ( 0 , k , \lambda ), %\label{eq:w(r,k,lambda)_1}
\end{align*}
and likewise for $\tilde w_{\ge \sigma}( 0 , k , \lambda)$ except that $h^A_{\ge\sigma}$, $h^B_{\ge\sigma}$ are replaced by $\bar{h}^A_{\ge\sigma}$, $\bar{h}^B_{\ge\sigma}$. (Observe in particular that the terms proportional to $(P-P_\ph)$ vanish. This is due to the fact that the charge of the total system we consider vanishes).

Moreover, we have that $\bar h^A_{\ge\sigma} ( 0 , k , \lambda ) = h^A_{\ge\sigma} ( 0 , k , \lambda )$ and
\begin{equation*}
\Pi_{0,\ge\sigma} p_r \cdot h^A_{\ge\sigma} ( 0 , k , \lambda ) ( \mathds{1} \otimes \pi_0 \otimes \mathds{1} ) = 0.
\end{equation*}
This yields
\begin{align*}
&\int_{\mathbb{R}^3} \Pi_{0,\ge\sigma} \tilde w_{\ge\sigma}(0,k,\lambda) \big [ H_r + \frac{1}{2M} ( P - k )^2 + |k| - E_{g,\ge\sigma}(P) \big ]^{-1} w_{\ge\sigma}(0,k,\lambda) \Pi_{0,\ge\sigma} \d k \notag \\
&= g^2 \big ( \tilde \Gamma^{A,\mathrm{diag}}_{\ge\sigma}(g,P,\lambda) + \tilde \Gamma^B_{\ge\sigma}(g,P,\lambda) \big ), \phantom{\int}
\end{align*}
where
\begin{align*}
\tilde \Gamma^{A,\mathrm{diag}}_{\ge\sigma}(g,P,\lambda) :=& \frac{1}{\mu^2} \int_{\mathbb{R}^3} \Pi_{0,\ge\sigma} p_r \cdot h^A_{\ge\sigma} ( 0 , k , \lambda ) \big [ H_r + \frac{1}{2M} ( P - k )^2 + |k| - E_{g,\ge\sigma}(P) \big ]^{-1} \notag \\
& \phantom{ \frac{1}{\mu^2} \int_{\mathbb{R}^3} } ( \mathds{1} \otimes \bar \pi_0 \otimes \mathds{1} ) p_r \cdot h^A_{\ge\sigma} ( 0 , k , \lambda ) \Pi_{0,\ge\sigma} \d k,
\end{align*}
and
\begin{align*}
\tilde \Gamma^B_{\ge\sigma}(g,P,\lambda): =& \int_{\mathbb{R}^3} \Pi_{0,\ge\sigma} \Big ( - \frac{1}{2m_\el} \sigma^{\el} \cdot \bar h^B_{\ge\sigma}( 0 , k , \lambda ) + \frac{1}{2m_\n} \sigma^{\n} \cdot \bar h^B_{\ge\sigma} ( 0 , k , \lambda ) \Big ) \notag \\
&\phantom{ \quad  \int_{\mathbb{R}^3} } \big [ e_0 + \frac{1}{2M} ( P - k )^2 + |k| - E_{g,\ge\sigma}(P) \big ]^{-1} \notag \\
& \phantom{ \quad \int_{\mathbb{R}^3} } \Big ( - \frac{1}{2m_\el} \sigma^{\el} \cdot h^B_{\ge\sigma}( 0 , k , \lambda ) + \frac{1}{2m_\n} \sigma^{\n} \cdot h^B_{\ge\sigma} ( 0 , k , \lambda ) \Big ) \Pi_{0,\ge\sigma} \d k.
\end{align*}
For any $|P| \le M/2$ and $|k| \ge \sigma$, we have that
\begin{align*}
\Big \| \big [ H_r + \frac{1}{2M} ( P - k )^2 + |k| - E_{0}(P) \big ]^{-1} ( \mathds{1} \otimes \bar \pi_0 \otimes \mathds{1} ) \Big \| \le \frac{1}{e_1-e_0},
\end{align*}
and hence also that 
\begin{align*}
\Big \| \big [ H_r + \frac{1}{2M} ( P - k )^2 + |k| - E_{g,\ge\sigma}(P) \big ]^{-1} ( \mathds{1} \otimes \bar \pi_0 \otimes \mathds{1} ) \Big \| \le \frac{1}{e_1-e_0},
\end{align*}
by Lemma \ref{lm:Eg-E0}. Therefore, using the first resolvent equation together with the facts that $|E_0(P) - E_{g,\ge\sigma}(P)| \lesssim g^2$ (see Lemma \ref{lm:Eg-E0}) and $| h^A_{\ge\sigma}(0,k,\lambda) | \lesssim \mathds{1}_{|k|\ge\sigma}(k) |k|^{-1/2} \chi_\Lambda(k)$, we get
\begin{align*}
\tilde \Gamma^{A,\mathrm{diag}}_{\ge\sigma}(g,P,\lambda) = \Gamma^{A,\mathrm{diag}}_{\ge\sigma}(P,\lambda) + \mathrm{Rem}^A(g,P,\sigma),
\end{align*}
where
\begin{align*}
\| \mathrm{Rem}^A(g,P,\sigma) \| \lesssim g^2 \int_{ \mathbb{R}^3 } \mathds{1}_{ \sigma \le |k| \le \Lambda } \frac{ \d k }{ |k| } \lesssim g^2.
\end{align*}
Likewise, for any $|P| \le M/2$ and $|k| \ge \sigma$, we have that
\begin{align*}
\big | \big [ e_0 + \frac{1}{2M} ( P - k )^2 + |k| - E_{g,\ge\sigma}(P) \big ]^{-1} \big | \le \frac{ 2 }{ |k| },
\end{align*}
and since $| h^B_{\ge\sigma}(0,k,\lambda) | \lesssim \mathds{1}_{|k|\ge\sigma}(k) |k|^{1/2} \chi_\Lambda(k)$, we thus obtain that
\begin{align*}
\tilde \Gamma^B_{\ge\sigma}(g,P,\lambda) = \Gamma^B_{\ge\sigma}(P,\lambda) + \mathrm{Rem}^B(g,P,\sigma),
\end{align*}
where
\begin{align*}
\| \mathrm{Rem}^B(g,P,\sigma) \| \lesssim g^2 \int_{ \mathbb{R}^3 } \mathds{1}_{ \sigma \le |k| \le \Lambda } \frac{ \d k }{ |k| } \lesssim g^2.
\end{align*}
Hence the lemma is proven.
\end{proof}
To conclude this subsection, we estimate the size of the splitting induced by the second order term in \eqref{eq:a1}. Since the matrix $\Gamma^{A,\mathrm{diag}}_{\ge\sigma}(P,\lambda)$ of Lemma \ref{lm:large5} is diagonal, only $\Gamma^B_{\ge\sigma}(P,\lambda)$ is responsible for this splitting.

From now on, to simplify a few computations and since the system is rotation invariant, we choose the total momental $P$ to be directed along $\overrightarrow{ \epsilon_3 }$.
\begin{lemma}\label{lm:large6}
There exist $p_c>0$ and $\sigma_c > 0$ such that, for all $P = P_3 \overrightarrow {\epsilon_3}$ satisfying $0 \le |P| \le p_c$, for all $0 < \sigma \le \sigma_c$, and $\lambda \in \{1,2\}$,
\begin{align*}
& \Gamma^B_{\ge\sigma}(P,\lambda) = \Gamma^{B,\mathrm{diag}}_{\ge\sigma}(P,\lambda) + \Gamma^{B,\#}_{\ge\sigma}(P,\lambda),
\end{align*}
where $\Gamma^{B,\mathrm{diag}}_{\ge\sigma}(P,\lambda)$ is the diagonal operator on $\mathrm{Ran} \, \Pi_{0,\ge\sigma}$ given by
\begin{align*}
\Gamma^{B,\mathrm{diag}}_{\ge\sigma}(P,\lambda) := \Big ( \frac{1}{4m_\el^2} + \frac{1}{4m_\n^2} \Big ) \int_{\mathbb{R}^3} \frac{ \big | h^B_{\ge\sigma}( 0 , k , \lambda ) \big |^2 \d k }{ |k| - ( k \cdot P ) / M + k^2/2M } \Pi_{0,\ge\sigma},
\end{align*}
and
\begin{align*}
\Gamma^{B,\#}_{\ge\sigma}(P,\lambda) := - \frac{1}{2m_\el m_\n} \sum_{j=1,2,3} \int_{\mathbb{R}^3} \frac{ \big | h^B_{j,\ge\sigma}( 0 , k , \lambda ) \big |^2 \d k }{ |k| - ( k \cdot P ) / M + k^2/2M } \Pi_{0,\ge\sigma} \sigma^\el_j \sigma^\n_j \Pi_{0,\ge\sigma} .
\end{align*}
\end{lemma}
\begin{proof}
Using standard properties of the Pauli matrices (see Lemma \ref{lm:Pauli1}), since for any $k\in\mathbb{R}^3$ and $\lambda \in \{1,2\}$, $\bar h^B_{\ge\sigma}( 0 , k , \lambda ) = - h^B_{\ge\sigma}( 0 , k , \lambda )$, we have that
\begin{align*}
\big ( \sigma^{\el} \cdot \bar h^B_{\ge\sigma}( 0 , k , \lambda ) \big ) \big ( \sigma^{\el} \cdot h^B_{\ge\sigma}( 0 , k , \lambda ) \big ) = - ( h^B_{\ge\sigma}( 0 , k , \lambda ) )^2 = \big | h^B_{\ge\sigma}( 0 , k , \lambda ) \big |^2,
\end{align*}
and likewise with $\sigma^\n$ replacing $\sigma^\el$. Next, we observe that for $P = P_3 \overrightarrow{\epsilon_3}$, $\lambda \in \{1,2\}$ and $j,j' \in \{1,2,3\}$, $j\neq j'$,
\begin{align*}
\int_{ \mathbb{R}^3 } \frac{ \bar h^B_{j,\ge\sigma}( 0 , k , \lambda ) h^B_{j',\ge\sigma}( 0 , k , \lambda ) }{ |k| - ( k \cdot P ) / M + k^2/2M } \d k =0.
\end{align*}
The lemma then follows straightforwardly from the expression of $\Gamma^B_{\ge\sigma}(P,\lambda)$ given in the statement of Lemma \ref{lm:large5}.
\end{proof}
\begin{lemma}\label{lm:large7}
There exist $p_c>0$ and $\sigma_c > 0$ such that, for all $P = P_3 \overrightarrow {\epsilon_3}$ satisfying $0 \le |P| \le p_c$ and for all $0 < \sigma \le \sigma_c$, the eigenvalues of the operator
\begin{align*}
\Gamma^{B,\#}_{\ge\sigma}(P) := - \sum_{\lambda=1,2} \Gamma^{B,\#}_{\ge\sigma}(P,\lambda) 
\end{align*}
are given by
\begin{align*}
& \gamma^{(0)}_{\ge\sigma}(P) := - \frac{1}{8 m_\el m_\n \pi^2} \int_{ \mathbb{R}^3 } \frac{ |k| \mathds{1}_{\sigma \le |k| \le \Lambda}(k) }{ |k| - ( k \cdot P ) / M + k^2/2M } \d k, \notag \\
& \gamma^{(j)}_{\ge\sigma}(P) := \frac{1}{8 m_\el m_\n \pi^2} \int_{ \mathbb{R}^3 } \frac{ |k| \mathds{1}_{\sigma \le |k| \le \Lambda}(k) }{ |k| - ( k \cdot P ) / M + k^2/2M } \frac{ k_j^2 }{ |k|^2 } \d k,  \quad j=1,2,3.
\end{align*}
\end{lemma}
\begin{proof}
It directly follows from the properties of the Pauli matrices (see Lemma \ref{lm:Pauli2}).
\end{proof}
\begin{remarks} \label{rk:gap}
$\quad$
\begin{enumerate}
\item For $P=0$, we observe that $\gamma^{(1)}_{\ge\sigma}(0)=\gamma^{(2)}_{\ge\sigma}(0)=\gamma^{(3)}_{\ge\sigma}(0)$. It may however not be the case for $P \neq 0$.
\item \label{item:gapii} The gap above the lowest eigenvalue $\gamma^{(0)}_{\ge\sigma}(P)$ is non-vanishing. More precisely, letting
\begin{align*}
\delta_{\ge\sigma}(P) := \min \left ( \gamma^{(1)}_{\ge\sigma}(P) , \gamma^{(2)}_{\ge\sigma}(P) , \gamma^{(3)}_{\ge\sigma}(P) \right ) - \gamma^{(0)}_{\ge\sigma}(P),
\end{align*}
and
\begin{align*}
\delta := \inf_{0\le |P| \le p_c , 0 \le \sigma \le \Lambda/2 } \delta_{\ge\sigma}(P),
\end{align*}
we have
\begin{align*}
\delta \ge  \frac{1}{8 m_\el m_\n \pi^2} \int_{ \mathbb{R}^3 } \frac{ |k| \mathds{1}_{\Lambda / 2 \le |k| \le \Lambda}(k) }{ ( 1 + p_c / M ) |k| + k^2 / 2M} \d k > 0.
\end{align*}
\end{enumerate}
\end{remarks}

\subsection{Proof of Theorem \ref{thm:large_sigma}}
$\quad$

\noindent \emph{Proof of Theorem \ref{thm:large_sigma}}.
Let $p_c$ be fixed as the minimum of the $p_c$'s given by Lemmas \ref{lm:term_order_g^2}--\ref{lm:large7} and \ref{lm:Eg-E0}, and let $\mathrm{C}_W$ be given by Lemma \ref{lm:estimate_Wg}. Let $\sigma_c = \beta_{c2} = \varepsilon$ where $\varepsilon>0$ is a small, fixed parameter (smaller, in particular, than the minimum of the $\sigma_c$'s given by Lemmas \ref{lm:term_order_g^2_lm1}--\ref{lm:large7} and than the $\beta_{c2}$'s given by Lemma \ref{lm:term_order_g^2}). Let now $0 < \beta_{c1} < \beta_{c2}$ and let $g_c$ be fixed smaller than the minimum of the $g_c$'s given by Lemmas \ref{lm:term_order_g^2}--\ref{lm:large7}. We recall in particular from the proofs of Lemma \ref{lm:term_order_g^2} that $g_c^2 \le \beta_{c1} \sigma_c$, which implies that for all $0 < g \le g_c$ and $\beta \ge \beta_{c1}$, we have that $0 < \sigma_{g,\beta} \le \sigma_c$, with $\sigma_{g,\beta} = g^2 \beta^{-1}$.

Let $0 \le |P| \le p_c$, $0< g \le g_c$, $\beta_{c1} \le \beta \le \beta_{c2}$. By rotation invariance, we can assume that $P = P_3 \overrightarrow{\epsilon_3}$.
Before starting the proof we introduce a few more notations to simplify expressions. Combining Lemmas \ref{lm:term_order_g^2}--\ref{lm:large6}, we can write
\begin{align*}
F_{g,\ge\sigma}( P ) =& ( E_0(P) - E_{g,\ge\sigma}(P)) \Pi_{0,\ge\sigma} + g^2 \mathrm{d}_{\ge \sigma }(P) \Pi_{0,\ge\sigma} + g^2 \Gamma^{B,\#}_{\ge\sigma}(P) + \mathrm{Rem} ( g , P , \beta ), 
\end{align*}
where $\mathrm{d}_{\ge \sigma }(P)$ is the (diagonal) bounded operator on $\mathrm{Ran} ( \Pi_{0,\ge\sigma} )$ defined by
\begin{align*}
\mathrm{d}_{\ge \sigma }(P) :=& - \sum_{ \lambda = 1 , 2 } \Big ( \Gamma^{A,\mathrm{diag}}_{\ge\sigma}( P , \lambda ) + \Gamma^{B,\mathrm{diag}}_{\ge\sigma}(P,\lambda) \Big ),
\end{align*}
and $\mathrm{Rem}( g , P , \beta )$ is a bounded operator on $\mathrm{Ran} ( \Pi_{0,\ge\sigma} )$ satisfying
\begin{align}
\| \mathrm{Rem} ( g , P , \beta ) \| &\lesssim g^2 ( \beta^{ \frac{1}{2} } + g + 2 g^{ \frac{2}{3} } + 2 g^2 ) \lesssim g^2 \varepsilon^{ \frac{1}{2} }.  \label{eq:estimate_R()}
\end{align}
Here we used that, by assumption, $\beta$ and $g$ are smaller than $\varepsilon$. We also introduce the operator
\begin{align*}
F^{(2)} :=& F_{g,\ge\sigma}(P) + E_{g,\ge\sigma}(P) \Pi_{0,\ge\sigma} - \mathrm{Rem} ( g , P , \beta ) \notag \\
&= \big ( E_0(P) + g^2 \mathrm{d}_{\ge \sigma }(P) \big ) \Pi_{0,\ge\sigma} + g^2 \Gamma^{B,\#}_{\ge\sigma}(P).
\end{align*}
Identifying $\mathrm{d}_{\ge\sigma}(P)$ with a scalar quantity, the lowest eigenvalue of $F^{(2)}$ is, according to Lemma \ref{lm:large7}, given by
\begin{align*}
e^{(0)}_{\ge \sigma}(P) := E_0(P) + g^2 ( \mathrm{d}_{\ge \sigma }(P) + \gamma^{(0)}_{\ge\sigma}(P) ).
\end{align*}
Moreover, it follows again from Lemma \ref{lm:large7} that $e^{(0)}_{\ge \sigma}(P)$ is simple and that
\begin{align}\label{eq:Gap_tildeF2}
\mathrm{Gap} ( F^{(2)} ) \ge g^2 \delta_{\ge\sigma}(P) \ge g^2 \delta, 
\end{align}
on $\mathrm{Ran}( \Pi_{0,\ge\sigma} )$, where $\delta_{\ge\sigma}(P)$  and $\delta>0$ are given by Remark \ref{rk:gap} (\ref{item:gapii}). 

Let $\phi^{(0)}_{\ge \sigma}(P) \in \mathrm{Ran}( \Pi_{0,\ge\sigma} )$ denote a normalized eigenstate associated with the eigenvalue $e^{(0)}_{\ge \sigma}(P)$ of $F^{(2)}$. \\

\noindent \textbf{Step 1}\quad Let us prove that 
\begin{align}\label{eq:ineqEg}
e^{(0)}_{\ge \sigma}(P) \ge E_{g,\ge\sigma}(P) - \frac{ g^2 \delta }{ 4 }.
\end{align}
Let $\psi^{(0)}_{\ge \sigma}(P)$ be the following trial state:
\begin{align*}
\psi^{(0)}_{\ge \sigma}(P) := \phi^{(0)}_{\ge \sigma}(P) \, - & \big [ K_{0,\ge\sigma}(P) - E_{g,\ge\sigma}(P) + \bar \Pi_{0,\ge\sigma} W_{g,\ge\sigma}(P) \bar \Pi_{0,\ge\sigma} \big ]^{-1} \notag \\
& \quad \bar \Pi_{0,\ge\sigma} W_{g,\ge\sigma}(P) \phi^{(0)}_{\ge \sigma}(P).
\end{align*}
We observe that $\langle \psi^{(0)}_{\ge \sigma}(P) , \phi^{(0)}_{\ge \sigma}(P) \rangle = \langle \phi^{(0)}_{\ge \sigma}(P) , \phi^{(0)}_{\ge \sigma}(P) \rangle = 1$. Proceeding as in the proof of Lemma \ref{lm:term_order_g^2}, using \eqref{eq:H>F} and \eqref{eq:Neumann_estimate_3}, we find that
\begin{align}
\| \psi^{(0)}_{\ge \sigma}(P) - \phi^{(0)}_{\ge \sigma}(P) \| &\le 4 \sqrt{2} \mathrm{C}_W \beta^{\frac{1}{2}} \lesssim \varepsilon^{\frac 12}, \label{eq:||tildepsi||}
\end{align}
where we used that $\beta^{ 1/2 } \le \varepsilon^{ 1/2 }$. Moreover, using the properties of the Feshbach operator, we obtain that
\begin{align*}
& \Pi_{0,\ge\sigma} \big ( K_{g,\ge \sigma}(P) - E_{g,\ge\sigma}(P) \big ) \psi^{(0)}_{\ge \sigma}(P) \notag \\
& = F_{g,\ge\sigma}(P) \phi^{(0)}_{\ge \sigma}(P) \notag \\
& = \big ( e^{(0)}_{\ge \sigma}(P) -  E_{g,\ge\sigma}(P) \big ) \phi^{(0)}_{\ge \sigma}(P) + \mathrm{Rem}(g,P,\beta) \phi^{(0)}_{\ge \sigma}(P),
\end{align*}
and
\begin{align*}
& \bar \Pi_{0,\ge\sigma} \big ( K_{g,\ge \sigma}(P) - E_{g,\ge\sigma}(P) \big ) \psi^{(0)}_{\ge \sigma}(P) \notag \\
& = \bar \Pi_{0,\ge\sigma} W_{g,\ge\sigma}(P) \phi^{(0)}_{\ge \sigma}(P) - \bar \Pi_{0,\ge\sigma} W_{g,\ge\sigma}(P) \phi^{(0)}_{\ge \sigma}(P) = 0.
\end{align*}
Therefore we deduce that
\begin{align*}
0 & \le \big \langle \psi^{(0)}_{\ge \sigma}(P) , \big ( K_{g,\ge \sigma}(P) - E_{g,\ge\sigma}(P) \big ) \psi^{(0)}_{\ge \sigma}(P) \big \rangle \notag \\
& = e^{(0)}_{\ge \sigma}(P) - E_{g,\ge\sigma}(P) + \langle \psi^{(0)}_{\ge \sigma}(P) , \mathrm{Rem}(g,P,\beta) \phi^{(0)}_{\ge \sigma}(P) \rangle \notag \\
& \le e^{(0)}_{\ge \sigma}(P) - E_{g,\ge\sigma}(P) + \| \psi^{(0)}_{\ge \sigma}(P) \| \|\mathrm{Rem}(g,P,\beta) \| \notag \\
& \le e^{(0)}_{\ge \sigma}(P) - E_{g,\ge\sigma}(P) + \mathrm{C} g^2 \varepsilon^{\frac{1}{2}},
\end{align*}
where in the last inequality we used \eqref{eq:||tildepsi||} and \eqref{eq:estimate_R()}. Hence, choosing $\varepsilon^{1/2} \ll \delta$, \eqref{eq:ineqEg} is proven. \\

\noindent \textbf{Step 2}\quad Let $\mu_2$ denote the second point above $E_{g,\ge\sigma}(P)$ in the spectrum of $K_{g,\ge\sigma}(P)$. The min-max principle implies that
\begin{align*}
\mu_2 \ge \underset{ \begin{small} \begin{array}{c} \psi \in D( K_{g,\ge\sigma}( P ) ) , \| \psi \|=1 , \\ \psi \in [ \phi^{(0)}_{\ge \sigma}(P) ]^\perp \end{array} \end{small} }{ \inf } \langle  \psi , K_{g,\ge\sigma}(P) \psi \rangle.
\end{align*}
Now, for $\psi$ as above, using inequality \eqref{eq:ineq_Fesh} of Lemma \ref{lm:term_order_g^2}, we can write
\begin{align*}
\langle \psi , K_{g,\ge\sigma}(P) \psi \rangle & \ge \langle \psi , F_{g,\ge\sigma}(P) \psi \rangle + E_{g,\ge\sigma}(P) \notag \\
& \ge \langle \psi , F^{(2)} \psi \rangle + \langle \psi , \mathrm{Rem}(g,P,\beta) \psi \rangle.
\end{align*}
Since $F^{(2)} \bar{\Pi}_{0,\ge\sigma} = 0$ and since $\psi \in [ \phi^{(0)}_{\ge \sigma}(P) ]^\perp$, it follows from \eqref{eq:Gap_tildeF2} that
\begin{align*}
\langle \psi , F^{(2)} \psi \rangle \ge ( e^{(0)}_{\ge \sigma}(P) + g^2 \delta ) \| \Pi_{0,\ge\sigma} \psi \|^2 \ge e^{(0)}_{\ge \sigma}(P) + g^2 \delta.
\end{align*}
In the last inequality, we used that $E_0(P) = P^2/2M + e_0 = P^2/2M - \mu / 2$ is negative for $|P| \le p_c$ small enough, and hence that $e^{(0)}_{\ge \sigma}(P) + g^2 \delta$ is also negative for $g$ and $|P|$ small enough. 
Moreover, \eqref{eq:estimate_R()} yields
\begin{align*}
\langle \psi , \mathrm{Rem}(g,P,\beta) \psi \rangle \ge - \mathrm{C} g^2 \varepsilon^{\frac 12}.
\end{align*}
Thus, applying Step 1, we obtain that
\begin{align*}
\mu_2 \ge E_{g,\ge\sigma}(P) + \frac{3}{4} g^2 \delta - \mathrm{C} g^2 \varepsilon^{\frac 12} \ge E_{g,\ge\sigma}(P) + \frac{1}{2} g^2 \delta,
\end{align*}
provided $\varepsilon$ is small enough. Changing notations ($\delta / 2 \to \delta$), this implies the statement of the theorem. \qed

%%%%%%%%%%%%%%%%%%%%%%%%%%%%%%%%%%%%%%%%%%
%%%%%%%%%%%%%%%%%%%%%%%%%%%%%%%%%%%%%%%%%%
%%%%%%%%%%%%%%%%%%%%%%%%%%%%%%%%%%%%%%%%%%
%%%%%%%%%%%%%%%%%%%%%%%%%%%%%%%%%%%%%%%%%%
\section{Proof of Theorem \ref{thm:main}}\label{section:3}

The main theorem of this section, which will allow us to prove Theorem \ref{thm:main}, is the following.
\begin{theorem}\label{thm:induction}
There exist $g_c>0$, $p_c>0$ and $0 < \eta \le1/4$ such that, for all $0 < g \le g_c$ and $0 \le |P| \le p_c$, there exist $\sigma_c > 0$ such that for all $0<\sigma\le\sigma_c$,
\begin{align*}
\mathbf{Gap}( g , P , \sigma , \eta ) \text{ holds}. 
\end{align*}
\end{theorem}
We need the following two lemmas before starting the proof of Theorem \ref{thm:induction}.
\begin{lemma}\label{lm:gap1}
Let $g_c>0$, $\sigma_c>0$ and $p_c>0$ be fixed small enough and let $0 \le g \le g_c$, $0 \le \sigma \le \sigma_c$ and $0 \le |P| \le p_c$ be such that $\mathbf{Gap}( g , P , \sigma , \eta )$ holds. Let $\Phi_{g,\ge\sigma}( P )$ be a normalized ground state of $K_{g,\ge\sigma} ( P )$. For all $0 \le \tau \le \sigma$, $E_{g,\ge\sigma}(P)$ is a simple eigenvalue of $H_{g,\ge\sigma}( P ) |_{\mathcal{H}_{\fib,\ge\tau}}$ associated with the normalized ground state $\Phi_{g,\ge\sigma}(P) \otimes \Omega_{\ge\tau}^{\le\sigma}$, and
\begin{equation*}
\mathrm{Gap} ( H_{g,\ge\sigma}( P ) |_{\mathcal{H}_{\fib,\ge\tau}} ) \ge \min ( \eta \sigma , \frac{ \tau }{ 4 } ).
\end{equation*}
\end{lemma}
\begin{proof}
First, one can readily verifies that $\Phi_{g,\ge\sigma}(P) \otimes \Omega_{\ge\tau}^{\le\sigma}$ is an eigenstate of $H_{g,\ge\sigma}(P) |_{\mathcal{H}_{\fib,\ge\tau}}  $ associated with the eigenvalue $E_{g,\ge\sigma}(P)$. Since
\begin{align*}
\mathds{1} - \big ( \Pi_{g,\ge\sigma}( P ) \otimes \Pi_{ \Omega_{\ge\tau}^{\le\sigma} } \big ) = \big ( \mathds{1} - \Pi_{g,\ge\sigma}( P ) \big ) \otimes \Pi_{ \Omega_{\ge\tau}^{\le\sigma} } + \mathds{1} \otimes \big ( \mathds{1} - \Pi_{ \Omega_{\ge\tau}^{\le\sigma} } \big ),
\end{align*}
and since $\mathds{1} \otimes \Pi_{ \Omega_{\ge\tau}^{\le\sigma} }$ commutes with $H_{g,\ge\sigma}( P ) |_{\mathcal{H}_{\fib,\ge\tau}}$, we have that
\begin{align}
& \underset{ \begin{small} \begin{array}{c} \psi \in D( H_{g,\ge\sigma}( P ) |_{\mathcal{H}_{\fib,\ge\tau}} ) , \| \psi \|=1 , \\ \psi \in [ \Phi_{g,\ge\sigma}(P) \otimes \Omega_{\ge\tau}^{\le\sigma}]^\perp \end{array} \end{small} }{ \inf } \langle  \psi , H_{g,\ge\sigma}(P) |_{\mathcal{H}_{\fib,\ge\tau}} \psi \rangle \notag \\
& \ge \min \Big ( \underset{ \begin{small} \begin{array}{c} \psi \in D( H_{g,\ge\sigma}( P ) |_{\mathcal{H}_{\fib,\ge\tau}} ) , \| \psi \|=1 , \\ \psi \in [ \Phi_{g,\ge\sigma}(P) ]^\perp \otimes [\Omega_{\ge\tau}^{\le\sigma}] \end{array} \end{small} }{ \inf } \langle \psi , H_{g,\ge\sigma}(P) |_{\mathcal{H}_{\fib,\ge\tau}} \psi \rangle , \notag \\
& \phantom{ \ge \min \Big ( } \underset{ \begin{small} \begin{array}{c} \psi \in D( H_{g,\ge\sigma}( P ) |_{\mathcal{H}_{\fib,\ge\tau}} ) , \| \psi \|=1 , \\ \psi \in \mathcal{H}_{\fib,\ge\sigma} \otimes [\Omega_{\ge\tau}^{\le\sigma}]^\perp \end{array} \end{small} }{ \inf } \langle \psi , H_{g,\ge\sigma}(P) |_{\mathcal{H}_{\fib,\ge\tau}} \psi \rangle \Big ). \label{eq:min(inf,inf)}
\end{align}
The assumption that $\mathrm{Gap}( K_{g,\ge\sigma}(P) ) \ge \eta \sigma$ yields that
\begin{align*}
\underset{ \begin{small} \begin{array}{c} \psi \in D( H_{g,\ge\sigma}( P ) |_{\mathcal{H}_{\fib,\ge\tau}} ) , \| \psi \|=1 , \\ \psi \in [ \Phi_{g,\ge\sigma}(P) ]^\perp \otimes [\Omega_{\ge\tau}^{\le\sigma}] \end{array} \end{small} }{ \inf } \langle \psi , H_{g,\ge\sigma}(P) |_{\mathcal{H}_{\fib,\ge\tau}} \psi \rangle \ge E_{g,\ge\sigma}(P) + \eta \sigma.
\end{align*}
To estimate from below the second term in the right-hand-side of \eqref{eq:min(inf,inf)}, we use the fact that $H_{g,\ge\sigma}(P) |_{\mathcal{H}_{\fib,\ge\tau}}$ commutes with the number operator $\mathds{1} \otimes \mathcal{N}_{\ph,\ge\tau}^{\le\sigma}$ (defined in \eqref{eq:defNtausigma}). We then consider $\psi \in D( H_{g,\ge\sigma}( P ) |_{\mathcal{H}_{\fib,\ge\tau}} )$ such that $\| \psi \|=1$ and $\mathds{1} \otimes \mathcal{N}_{\ph,\ge\tau}^{\le\sigma} \psi = n \psi$ with $n \ge 1$. This state $\psi$ may be seen as an element of $\mathrm{L}^2( \{ (k,\lambda), \tau \le |k| \le \sigma \} ; \mathcal{H}_{\mathrm{fib},\ge\sigma} \otimes \mathrm{L}^2( \{ (k,\lambda), \tau \le |k| \le \sigma \}^{n-1} ) )$, and we have that
\begin{align*}
& \langle \psi , H_{g,\ge\sigma}(P) |_{\mathcal{H}_{\fib,\ge\tau}} \psi \rangle \notag \\
&= \sum_{\lambda=1,2} \int_{ \tau \le |k| \le \sigma } \big \langle \big ( \psi(k,\lambda) , H_{g,\ge\sigma}(P - k) |_{\mathcal{H}_{\fib,\ge\sigma} \otimes \mathrm{L}^2( \{ (k,\lambda), \tau \le |k| \le \sigma \}^{n-1} ) } + |k| \big ) \psi(k,\lambda) \big \rangle \d k \notag \\
& \ge \inf_{\tau \le |k| \le \sigma}  \big ( E_{g,\ge\sigma}(P - k) + |k| \big ) \ge E_{g,\ge\sigma}(P) + \frac{ \tau }{ 4 },
\end{align*}
where the last inequality is a consequence of \cite[Lemma 4.3]{AGG2}. Therefore we obtain that
\begin{align*}
& \underset{ \begin{small} \begin{array}{c} \psi \in D( H_{g,\ge\sigma}( P ) |_{\mathcal{H}_{\fib,\ge\tau}} ) , \| \psi \|=1 , \\ \psi \in \mathcal{H}_{\fib,\ge\sigma} \otimes [\Omega_{\ge\tau}^{\le\sigma}]^\perp \end{array} \end{small} }{ \inf } \langle \psi , H_{g,\ge\sigma}(P) |_{\mathcal{H}_{\fib,\ge\tau}} \psi \rangle \ge E_{g,\ge\sigma}(P) + \frac{ \tau }{ 4 }, 
\end{align*}
which concludes the proof of the lemma.
\end{proof}
\begin{lemma}\label{lm:induction}
There exists $p_c>0$ such that, for all $\eta$ and $\kappa$ such that $0 < 4 \eta \le \kappa < 1$, there exists $g_c>0$ such that, for all $0 \le g \le g_c$, $0 \le |P| \le p_c$ and $\sigma>0$
\begin{equation*}
\mathbf{Gap}(g,P,\sigma,\eta) \Rightarrow \mathbf{Gap}(g,P,\kappa\sigma,\eta).
\end{equation*}
\end{lemma}
\begin{proof}
Assume that $\mathbf{Gap}(g,P,\sigma,\eta)$ is satisfied for some $\sigma > 0$. Let $\Phi_{g,\ge\sigma}(P)$ be a ground state of $K_{g,\ge\sigma}(P)$. Let $\tau = \kappa \sigma$ and let $\mu_2$ denote the first point above $E_{g,\ge\tau}(P)$ in the spectrum of $K_{g,\ge\tau}(P)$. By the min-max principle,
\begin{equation*}
\begin{split}
\mu_2 &\ge \underset{ \begin{small} \begin{array}{c} \psi \in D( K_{g,\ge\tau}(P) ) , \| \psi \| = 1 \\ \psi \in [ \Phi_{g,\ge\sigma}(P) \otimes \Omega_{\ge\tau}^{\le\sigma} ]^\perp \end{array} \end{small} }{ \inf } \langle \psi , K_{g,\ge\tau}(P) \psi \rangle,
\end{split}
\end{equation*}
where $[ \Phi_{g,\ge\sigma}(P) \otimes \Omega_{\ge\tau}^{\le\sigma} ]^\perp$ denotes the orthogonal complement of the vector space spanned by $\Phi_{g,\ge\sigma}(P) \otimes \Omega_{\ge\tau}^{\le\sigma}$ in $\mathcal{H}_{\fib,\ge\sigma} \otimes \mathcal{H}_{\ph,\ge\tau}^{\le\sigma} \simeq \mathcal{H}_{\fib,\ge\tau}$. It follows from Lemma \ref{lm:estimate_Wg2} that for any $\rho>0$,
\begin{align*}
\langle \psi , K_{g,\ge\tau}(P) \psi \rangle & = \langle \psi , H_{g,\ge\sigma}(P) |_{ \mathcal{H}_{\fib,\ge\tau} } \psi \rangle + \langle \psi , W_{g,\ge\tau}^{\le\sigma}(P) \psi \rangle \notag \\ 
&\ge \left [ 1 - \mathrm{C}_W g \sigma^{1/2} \rho^{-1/2} \right ] \langle \psi , H_{g,\ge\sigma}(P) |_{ \mathcal{H}_{\fib,\ge\tau} } \psi \rangle \notag \\
&\quad + \mathrm{C}_W g \sigma^{1/2} \rho^{-1/2} E_{g,\ge\sigma}(P) - \mathrm{C}_W g \sigma^{1/2} \rho^{1/2}.
\end{align*}
Next, by $\mathbf{Gap}(g,P,\sigma,\eta)$, Lemma \ref{lm:gap1} and the fact that $\tau = \kappa \sigma \ge 4 \eta \sigma$, we obtain that for any $\psi$ in $[ \Phi_{g,\ge\sigma}(P) \otimes \Omega_{\ge\tau}^{\le\sigma} ]^\perp$, $\| \psi \|=1$,
\begin{equation*}
\langle \psi , H_{g,\ge\sigma}(P) |_{ \mathcal{H}_{\fib,\ge\tau} } \psi \rangle \ge E_{g,\ge\sigma}(P) + \eta \sigma,
\end{equation*}
provided that $g$ is sufficiently small. Hence for any $\rho>0$ such that $\rho^{1/2} > \mathrm{C}_W g \sigma^{1/2}$,
\begin{equation*}
\begin{split}
\langle \psi , K_{g,\ge\tau}(P) \psi \rangle &\ge E_{g,\ge\sigma}(P) + \left [ 1 - \mathrm{C}_W g \sigma^{1/2} \rho^{-1/2} \right ] \eta \sigma - \mathrm{C}_W g \sigma^{1/2} \rho^{1/2}.
\end{split}
\end{equation*}
Choosing $\rho^{1/2} = 2 (1 - \kappa)^{-1} \mathrm{C}_W g \sigma^{1/2}$ and using Lemma \ref{lm:Eg-E02}, we get
\begin{align*}
\langle \psi , K_{g,\ge\tau}(P) \psi \rangle &\ge E_{g,\ge\sigma}(P) + \frac{1+\kappa}{2} \eta \sigma - 2 (1 - \kappa)^{-1} \mathrm{C}_W^2 g^2 \sigma \notag \\
& \ge E_{g,\ge\tau}(P) + \eta \tau + \frac{1-\kappa}{2} \eta \sigma - 2 (1 - \kappa)^{-1} \mathrm{C}_W^2 g^2 \sigma.
\end{align*}
Hence $\mu_2 \ge E_{g,\ge\tau}(P) + \eta\tau$ provided that $g^2 \le ( 4 \mathrm{C}_W^2 )^{-1} (1 - \kappa)^2 \eta$, which proves the lemma.
\end{proof}
 Let us now prove Theorem \ref{thm:induction}. \\

\noindent \emph{Proof of Theorem \ref{thm:induction}}.\quad
Let $p_c$ be the minimum of the $p_c$'s given by Theorem \ref{thm:large_sigma} and Lemma \ref{lm:induction}. Let $\beta_{c2}$ and $\delta$ be given by Theorem \ref{thm:large_sigma}. Note that, possibly by considering a smaller $\beta_{c2}$, we can assume that $4 \delta < \beta_{c2}^{-1}$. Fix $\beta_{c1}$ such that $0 < \beta_{c1} < \beta_{c2}$. Let $\eta := \delta \beta_{c1}$ and $\kappa := \beta_{c1} \beta_{c2}^{-1}$. In particular we have that $0 < 4 \eta \le \kappa < 1$. Let $g_c$ be the minimum of the $g_c$'s given by Theorem \ref{thm:large_sigma} and Lemma \ref{lm:induction}. 

Let $0 < g \le g_c$ and $0 \le |P| \le p_c$. By rotation invariance, we can assume that $P = P_3 \overrightarrow{ \epsilon_3 }$. Let $\sigma_c := g^2 \beta_{c1}^{-1}$. By Theorem \ref{thm:large_sigma}, we know that, for all $\sigma$ such that $g^2 \beta_{c2}^{-1} \le \sigma \le g^2 \beta_{c1}^{-1}$, $\mathbf{Gap}( g , P , \sigma , \eta )$ holds. Using Lemma \ref{lm:induction}, this implies that, for all $\sigma$ such that $g^2 \beta_{c2}^{-1} \le \sigma \le g^2 \beta_{c1}^{-1}$ and for all $n \in \mathbb{N}_0$, $\mathbf{Gap}( g , P , \kappa^n \sigma , \eta )$ holds. Since $\kappa = \beta_{c1} \beta_{c2}^{-1}$, we deduce that $\mathbf{Gap}( g , P , \sigma , \eta )$ is satisfied for all $0 < \sigma \le g^2 \beta_{c1}^{-1}$, which concludes the proof of the theorem. 
\\

We are now able to prove the main theorem of this paper. \\

\noindent \emph{Proof of Theorem \ref{thm:main}}.\quad
By \cite{AGG2}, we know that there exist $g_c>0$ and $p_c>0$ such that for all $0 \le g\le g_c$ and $0 \le |P| \le p_c$, $E_g(P)$ is an eigenvalue of $H_g(P)$. Let here $g_c>0$ and $p_c>0$ be fixed small enough (smaller, in particular, than the latter and than the ones given by Theorem \ref{thm:induction}). Let $0 < \eta \le 1/4$ be given by Theorem \ref{thm:induction} and let $\mathrm{C}_W$ be given by Lemma \ref{lm:estimate_Wg2}. Let $0 < g \le g_c$ and $0 \le |P| \le p_c$. By rotation invariance, we can assume without loss of generality that $P = P_3 \overrightarrow{ \epsilon_3 }$. Let $\sigma_c >0$ be given by Theorem \ref{thm:induction}.

For $0 < \sigma \le \sigma_c$, let $\Phi_{g,\ge\sigma}(P)$ denote a normalized ground state of $K_{g,\ge\sigma}(P)$. Recall from Lemma \ref{lm:gap1} that $\Phi_{g,\ge\sigma}(P) \otimes \Omega^{\le\sigma}$ is then a normalized ground state of $H_{g,\ge\sigma}(P)$. By the Banach-Alaoglu theorem, there exists a sequence $(\Phi_{g,\ge\sigma_n}(P) \otimes \Omega^{\le\sigma_n})_{n\in\mathbb{N}}$ which converges weakly as $n \to \infty$, with $0 < \sigma_n \le \mathrm{min}( \sigma_c , 1 )$ and $\sigma_n \to 0$. Let $\Phi_g(P)$ denote the corresponding limit. Using Lemma \ref{lm:barP0_Pg}, a standard argument (see for instance \cite{BFS,GLL}) shows that $\langle \Phi_g(P) , \Omega \rangle \neq 0$, from which it follows that $\Phi_g(P)$ is a normalized ground state of $H_g(P)$.

In order to prove Theorem \ref{thm:main}, it now suffices to follow \cite{BFP}. Assume by contradiction that there exists another normalized ground state, say $\Phi'_g(P)$, such that $\langle \Phi_g(P) , \Phi'_g(P) \rangle = 0$. We write
\begin{align}
\big | \langle \Phi_g(P) , \Phi'_g(P) \rangle \big |^2 & = \lim_{n \to \infty} \big | \langle \Phi_{g,\ge\sigma_n}(P) \otimes \Omega^{\le\sigma_n}, \Phi'_g(P) \rangle \big |^2 \notag \\
&= \lim_{n\to\infty} \big \langle \Phi'_g(P) , \mathds{1}_{ \{ E_{g,\ge\sigma_n}(P) \} }( H_{g,\ge\sigma_n}(P) ) \Phi'_g(P) \big \rangle \notag \\
&= 1 - \lim_{n\to\infty} \big \langle \Phi'_g(P) , ( \mathds{1} - \mathds{1}_{ \{ E_{g,\ge\sigma_n}(P) \} }( H_{g,\ge\sigma_n}(P) ) ) \Phi'_g(P) \big \rangle. \label{eq:PhigPhi'g}
\end{align}
Notice that, in the second equality, we used the fact that, by Theorem \ref{thm:induction} and Lemma \ref{lm:gap1},  $E_{g,\ge\sigma_n}(P)$ is a simple eigenvalue of $H_{g,\ge\sigma_n}(P)$. We decompose
\begin{align}
\mathds{1} - \mathds{1}_{ \{ E_{g,\ge\sigma_n}(P) \} }( H_{g,\ge\sigma_n}(P) ) =& (\mathds{1} - \mathds{1}_{ \{ E_{g,\ge\sigma_n}(P) \} }( K_{g,\ge\sigma_n}(P) ) ) \otimes \Pi_{ \Omega^{\le\sigma_n} } \notag \\
&+ \mathds{1} \otimes ( \mathds{1} - \Pi_{ \Omega^{\le\sigma_n} } ), \label{eq:decomp}
\end{align}
and use that, by Lemma \ref{lm:barP0_Pg_2},
\begin{align}
\big \langle \Phi'_g(P) , \mathds{1} \otimes ( \mathds{1} - \Pi_{ \Omega^{\le\sigma_n} } ) \Phi'_g(P) \big \rangle \le \big \langle \Phi'_g(P) , \mathcal{N}_\ph^{\le\sigma_n} \Phi'_g(P) \big \rangle \lesssim g^2 \sigma_n^2 \lesssim g^2. \label{eq:decomp_1}
\end{align}
On the other hand, by Theorem \ref{thm:induction}, we can write
\begin{align*}
& \big \langle \Phi'_g(P) , ( \mathds{1} - \mathds{1}_{ \{ E_{g,\ge\sigma_n}(P) \} }( K_{g,\ge\sigma_n}(P) ) ) \otimes \Pi_{ \Omega^{\le\sigma_n} } \Phi'_g(P) \big \rangle \phantom{ \frac{1}{\eta} } \notag \\
& \le \frac{1}{\eta\sigma_n}  \big \langle \Phi'_g(P) , ( K_{g,\ge\sigma_n}(P) - E_{g,\ge\sigma_n}(P) ) \otimes \Pi_{\Omega^{\le\sigma_n}} \Phi'_g(P) \big \rangle \notag \\
& = \frac{1}{\eta\sigma_n}  \big \langle \Phi'_g(P) , ( H_{g,\ge\sigma_n}(P) - E_{g,\ge\sigma_n}(P) ) ( \mathds{1} \otimes \Pi_{\Omega^{\le\sigma_n}} ) \Phi'_g(P) \big \rangle \notag \\
& \le \frac{1}{\eta\sigma_n}  \big \langle \Phi'_g(P) , ( H_{g,\ge\sigma_n}(P) - E_{g,\ge\sigma_n}(P) ) \Phi'_g(P) \big \rangle \notag \\
& = \frac{1}{\eta\sigma_n}  \big \langle \Phi'_g(P) , ( E_g(P) - E_{g,\ge\sigma_n}(P) - W_g^{\le\sigma_n}(P) ) \Phi'_g(P) \big \rangle,
\end{align*}
where in the last equality we used that $(H_g(P) - E_g(P)) \Phi'_g(P)=0$ and that $H_g(P) = H_{g,\ge\sigma_n}(P) + W_g^{\le\sigma_n}(P)$. By Lemmas \ref{lm:Eg-E02} and \ref{lm:estimate_Wg2}, this implies that
\begin{align}
& \big \langle \Phi'_g(P) , ( \mathds{1} - \mathds{1}_{ \{ E_{g,\ge\sigma_n}(P) \} }( K_{g,\ge\sigma_n}(P) ) ) \otimes \Pi_{ \Omega^{\le\sigma_n} } \Phi'_g(P) \big \rangle \le \frac{ \mathrm{C} g^2 \sigma_n }{ \eta } + \frac{ \mathrm{C}_W g }{ \eta } \lesssim g, \label{eq:decomp_2}
\end{align}
since $\sigma_n \le 1$. Combining \eqref{eq:decomp}, \eqref{eq:decomp_1} and \eqref{eq:decomp_2}, we obtain that, for $g$ small enough, $\eqref{eq:PhigPhi'g} > 0$, which is a contradiction. Hence the theorem is proven. \qed

%%%%%%%%%%%%%%%%%%%%%%%%%%%%%%%%%%%%%%%%%%%%%%%%%%%%%%%%%%%%%%%%%%%%%%%
%%%%%%%%%%%%%%%%%%%%%%%%%%%%%%%%%%%%%%%%%%%%%%%%%%%%%%%%%%%%%%%%%%%%%%%
%%%%%%%%%%%%%%%%%%%%%%%%%%%%%%%%%%%%%%%%%%%%%%%%%%%%%%%%%%%%%%%%%%%%%%%
%%%%%%%%%%%%%%%%%%%%%%%%%%%%%%%%%%%%%%%%%%%%%%%%%%%%%%%%%%%%%%%%%%%%%%%

%%%%%%%%%%%%%%%%%%%%%%%%%%%%%%%%%%%%%%%%%%%%%%%%%%%%%%%%
%%%%%%%%%%%%%%%%%%%%%%%%%%%%%%%%%%%%%%%%%%%%%%%%%%%%%%%%
%%%%%%%%%%%%%%%%%%%%%%%%%%%%%%%%%%%%%%%%%%%%%%%%%%%%%%%%
%%%%%%%%%%%%%%%%%%%%%%%%%%%%%%%%%%%%%%%%%%%%%%%%%%%%%%%%
\appendix

%%%%%%%%%%%%%%%%%%%%%%%%%%%%%%%%%%%%%%%%%%%%%%%%%%%%%%%%
%%%%%%%%%%%%%%%%%%%%%%%%%%%%%%%%%%%%%%%%%%%%%%%%%%%%%%%%
%%%%%%%%%%%%%%%%%%%%%%%%%%%%%%%%%%%%%%%%%%%%%%%%%%%%%%%%
%%%%%%%%%%%%%%%%%%%%%%%%%%%%%%%%%%%%%%%%%%%%%%%%%%%%%%%%
\section{Standard estimates}\label{section:estimates}

This appendix contains several fairly standard estimates which were used above. 
\begin{lemma}\label{lm:Hph<H_0}
For all $0 \le |P| < M$, and $\sigma \ge 0$, we have that
\begin{equation}\label{eq:Hph<H_0}
K_{0,\ge\sigma}(P) - E_0(P) \ge \big( 1 - \frac{ |P| }{ M } \big ) H_{\ph,\ge\sigma}.
\end{equation}
Moreover, the spectrum of $K_{0,\ge\sigma}(P)$ satisfies
\begin{align}\label{eq:Hph<H_0_1}
& \mathrm{spec}( K_{0,\ge\sigma}(P) ) \subseteq \big \{ E_0(P) \big \} \cup \big [ E_0(P) + \big ( 1 - \frac{ |P| }{ M } \big ) \sigma , + \infty ),
\end{align}
and
\begin{align}\label{eq:Hph<H_0_2}
& \mathrm{dim} \, \mathrm{Ker} \big ( K_{0,\ge\sigma}(P) - E_0(P) \big ) = 4.
\end{align}
\end{lemma}
\begin{proof}
Recall that
\begin{align*}
K_{0,\ge\sigma}(P) &= H_r + \frac{P^2}{2M} - \frac{1}{M} P \cdot P_{\ph,\ge\sigma} + \frac{1}{2M} P_{\ph,\ge\sigma}^2 + H_{\ph,\ge\sigma},
\end{align*}
on the Hilbert space $\mathcal{H}_{\fib,\ge\sigma}$. For $\Phi \in D( K_{0,\ge\sigma}(P) )$, we have that 
\begin{equation*}
| \langle \Phi , P \cdot P_{\ph,\ge\sigma} \Phi \rangle | \le |P| \langle \Phi , H_{\ph,\ge\sigma} \Phi \rangle.
\end{equation*}
Therefore,
\begin{align*}
K_{0,\ge\sigma}(P) &\ge e_0 + \frac{P^2}{2M} + \big ( 1 - \frac{ |P| }{ M } \big ) H_{\ph,\ge\sigma},
\end{align*}
which proves \eqref{eq:Hph<H_0}.

In order to prove \eqref{eq:Hph<H_0_1} and \eqref{eq:Hph<H_0_2}, it suffices to observe that
\begin{equation*}
K_{0,\ge\sigma}(P) ( y \otimes \phi_0 \otimes \Omega_{\ge\sigma} ) = E_0(P) ( y \otimes \phi_0 \otimes \Omega_{\ge\sigma} ),
\end{equation*}
for any $y \in \mathbb{C}^4$, and that, by \eqref{eq:Hph<H_0},
if $\psi \in D( K_{0,\ge\sigma}(P) )$ satisfies $\| \psi \| = 1$ and $\langle \psi , ( y \otimes \phi_0 \otimes \Omega_{\ge\sigma} ) \rangle = 0$ for all $y \in \mathbb{C}^4$, then
\begin{equation*}
\langle \psi , ( K_{0,\ge\sigma}(P) -E_0(P) ) \psi \rangle \ge \big ( 1 - \frac{ |P| }{ M } \big ) \sigma.
\end{equation*}
Hence the lemma follows.
\end{proof}
The number operator of particles of energies $\ge \sigma$ is defined by
\begin{equation*}
\mathcal{N}_{\ph,\ge\sigma} := \sum_{\lambda = 1,2} \int_{ |k| \ge \sigma } a^*_\lambda(k) a_\lambda(k) \d k.
\end{equation*}
The following two lemmas are easy generalizations of \cite[Lemma 4.5]{AGG2} based on the combination of a pull-through formula together with a Pauli-Fierz transformation (see also \cite{BFS,GLL}). We do not recall the proofs. 
\begin{lemma}\label{lm:barP0_Pg}
There exist $g_c>0$, $p_c>0$ and $\sigma_c>0$ such that, for all $0\le g\le g_c$, $0\le|P|\le p_c$ and $0 \le \sigma \le \sigma_c$, $\mathrm{Ker}( K_{g,\ge\sigma}(P) - E_{g,\ge\sigma}(P) ) \subset D( \mathcal{N}_{\ph,\ge\sigma}^{1/2})$, and the following holds: For all $\Phi_{g,\ge\sigma}(P) \in \mathrm{Ker}( K_{g,\ge\sigma}(P) - E_{g,\ge\sigma}(P) ) , \| \Phi_{g,\ge\sigma}(P) \| = 1$, we have that
\begin{align*}
\langle \Phi_{g,\ge\sigma}(P) , \mathcal{N}_{\ph,\ge\sigma} \Phi_{g,\ge\sigma}(P) \rangle \lesssim g^2.
\end{align*}
\end{lemma}
In the next lemma, we use the notation
\begin{align}
\mathcal{N}_{\ph,\ge\tau}^{\le\sigma} := \sum_{\lambda = 1,2} \int_{\tau \le |k| \le \sigma } a^*_\lambda(k) a_\lambda(k) \d k. \label{eq:defNtausigma}
\end{align}
\begin{lemma}\label{lm:barP0_Pg_2}
There exist $g_c>0$, $p_c>0$ and $\sigma_c>0$ such that, for all $0\le g\le g_c$, $0\le|P|\le p_c$ and $0 \le \tau \le \sigma \le \sigma_c$, $\mathrm{Ker}( K_{g,\ge\tau}(P) - E_{g,\ge\tau}(P) ) \subset D( ( \mathcal{N}_{\ph,\ge\tau}^{\le\sigma} )^{1/2})$, and the following holds: For all $\Phi_{g,\ge\tau}(P) \in \mathrm{Ker}( K_{g,\ge\tau}(P) - E_{g,\ge\tau}(P) ) , \| \Phi_{g,\ge\tau}(P) \| = 1$, we have that
\begin{align*}
\langle \Phi_{g,\ge\tau}(P) , \mathcal{N}_{\ph,\ge\tau}^{\le\sigma} \Phi_{g,\ge\tau}(P) \rangle \lesssim g^2 \sigma^2.
\end{align*}
\end{lemma}
The next lemma is proven in the same way as Lemma A.6 in \cite{AF2}. We do not reproduce the proof here.
\begin{lemma}\label{lm:Eg-E0}
There exist $g_c>0$, $p_c>0$, $\sigma_c>0$ and $\mathrm{C}_0>0$ such that, for all $0\le g\le g_c$, $0\le|P|\le p_c$, and $0 \le \sigma \le \sigma_c$,
\begin{equation*}
E_{g,\ge\sigma}(P) \le E_0(P) \le E_{g,\ge\sigma}(P) + \mathrm{C}_0 g^2.
\end{equation*}
\end{lemma}
Similarly we have the following:
\begin{lemma}\label{lm:Eg-E02}
There exist $g_c>0$, $p_c>0$, $\sigma_c>0$ and $\mathrm{C}_0>0$ such that, for all $0\le g\le g_c$, $0\le|P|\le p_c$, and $0 \le \tau \le \sigma \le \sigma_c$,
\begin{equation}\label{eq:Eg-E0_2}
E_{g,\ge\tau}(P) \le E_{g,\ge\sigma}(P) \le E_{g,\ge\tau}(P) + \mathrm{C}_0 g^2 \sigma^2.
\end{equation}
\end{lemma}
\begin{proof}
Again, the proof is similar to the one of Lemma A.6 in \cite{AF2}. We only sketch the main differences here. Recall that
\begin{align}\label{eq:Hgtausigma}
K_{g,\ge\tau}(P) = H_{g,\ge\sigma}(P) |_{ \mathcal{H}_{\fib,\ge\tau} } + W_{g,\ge\tau}^{\le\sigma}(P),
\end{align}
in $\mathcal{H}_{\mathrm{fib},\ge\tau}$, where $W_{g,\ge\tau}^{\le\sigma}(P)$ is given in \eqref{eq:Wg>tau<sigma}. By Lemma \ref{lm:gap1}, 
\begin{align*}
H_{g,\ge\sigma}(P) |_{ \mathcal{H}_{\fib,\ge\tau} } \big ( \Phi_{g,\ge\sigma}(P) \otimes \Omega_{\ge\tau}^{\le\sigma} \big ) = E_{g,\ge\sigma}(P) \Phi_{g,\ge\sigma}(P) \otimes \Omega_{\ge\tau}^{\le\sigma}.
\end{align*}
Thus, the first inequality in \eqref{eq:Eg-E0_2} follows from
\begin{align*}
E_{g,\ge\tau}(P) \le \big \langle \Phi_{g,\ge\sigma}(P) \otimes \Omega_{\ge\tau}^{\le\sigma} , K_{g,\ge\tau}(P) \Phi_{g,\ge\sigma}(P) \otimes \Omega_{\ge\tau}^{\le\sigma} \big \rangle = E_{g,\ge\sigma}(P),
\end{align*}
since $W_{g,\ge\tau}^{\le\sigma}(P)$ is Wick ordered.

In order to prove the second inequality in \eqref{eq:Eg-E0_2}, we use that, by \eqref{eq:Hgtausigma},
\begin{align*}
E_{g,\ge\sigma}(P) & \le \big \langle \Phi_{g,\ge\tau}(P) , H_{g,\ge\sigma}(P) |_{ \mathcal{H}_{\fib,\ge\tau} } \Phi_{g,\ge\tau}(P) \big \rangle \notag \\
& \le E_{g,\ge\tau}(P) - \big \langle \Phi_{g,\ge\tau}(P) , W_{g,\ge\tau}^{\le\sigma}(P) \Phi_{g,\ge\tau}(P) \big \rangle.
\end{align*}
Decomposing $W_{g,\ge\tau}^{\le\sigma}(P)$ into its expression \eqref{eq:Wg>tau<sigma} and estimating each term separately, we are able to conclude the proof thanks to Lemma \ref{lm:barP0_Pg_2}.
\end{proof}
The next two lemmas are proven in the same way as Lemma A.8 in \cite{AF2}.
\begin{lemma}\label{lm:estimate_Wg}
There exist $g_c>0$, $p_c>0$, $\sigma_c > 0$ and $\mathrm{C} _W> 0$ such that, for all $0 \le g \le g_c$, $0 \le |P| \le p_c$, $0 \le \sigma \le \sigma_c$ and $0 < \rho < 1$, the following holds:
\begin{align*}
& \big \| [ K_{0,\ge\sigma}(P) - E_0(P) + \rho ]^{-\frac{1}{2}} W_{g,\ge\sigma}(P) [ K_{0,\ge\sigma}(P) - E_0(P) + \rho ]^{-\frac{1}{2}} \big \| \le \mathrm{C}_W g \rho^{-\frac{1}{2}}. 
\end{align*}
\end{lemma}
\begin{lemma}\label{lm:estimate_Wg2}
There exist $g_c>0$, $p_c>0$, $\sigma_c > 0$ and $\mathrm{C} _W> 0$ such that, for all $0 \le g \le g_c$, $0 \le |P| \le p_c$, $0 \le \tau \le \sigma \le \sigma_c$ and $0 < \rho < 1$, the following holds:
\begin{align*}
& \big \| [ H_{g,\ge\sigma}(P) |_{ \mathcal{H}_{\fib,\ge\tau}} - E_{g,\ge\sigma}(P) + \rho ]^{-\frac{1}{2}} W_{g,\ge\tau}^{\le\sigma}(P) [ H_{g,\ge\sigma}(P) |_{ \mathcal{H}_{\fib,\ge\tau}} - E_{g,\ge\sigma}(P) + \rho ]^{-\frac{1}{2}} \big \| \notag \\
&\le \mathrm{C}_W g \sigma^{\frac{1}{2}} \rho^{-\frac{1}{2}}. 
\end{align*}
\end{lemma}
We conclude with two easy lemmas concerning Pauli matrices. Their proofs are left to the reader.
\begin{lemma}\label{lm:Pauli1}
Let $a = (a_1,a_2,a_3)$ denote a $3$-vector of complex numbers. We have that
\begin{equation*}
(a \cdot \sigma^\el) (a \cdot \sigma^\el ) = a_1^2 + a_2^2 + a_3^2 = ( a \cdot \sigma^\n ) (a \cdot \sigma^\n ).
\end{equation*}
\end{lemma}
\begin{lemma}\label{lm:Pauli2}
Let $a = (a_1,a_2,a_3)$ denote a $3$-vector of complex numbers. The eigenvalues of the $4\times 4$ matrix $a_1 \sigma^\el_1 \sigma^\n_1 + a_2 \sigma^\el_2 \sigma^\n_2 + a_3 \sigma^\el_3 \sigma^\n_3$ are the following:
\begin{equation*}
a_1 + a_2 - a_3,\quad a_1 - a_2 + a_3,\quad -a_1 + a_2 + a_3,\quad -a_1-a_2-a_3.
\end{equation*}
\end{lemma}

\bibliographystyle{amsplain}
\providecommand{\bysame}{\leavevmode\hbox to3em{\hrulefill}\thinspace}
\providecommand{\MR}{\relax\ifhmode\unskip\space\fi MR }
\providecommand{\MRhref}[2]{%
  \href{http://www.ams.org/mathscinet-getitem?mr=#1}{#2}
}
\providecommand{\href}[2]{#2}

\end{document}